\documentclass{aa}

\newcommand{\fdisc}{{f_{\rm disc}}}
\newcommand{\lcdm}{{$\Lambda$CDM}}

\usepackage[varg]{txfonts}

\usepackage{amssymb, aas_macros}
\usepackage{natbib}
\usepackage{color}
\usepackage{graphicx}

\begin{document}

\title{The Milky Way and Andromeda galaxies in a constrained \\
hydrodynamical simulation: morphological evolution }

\author{Cecilia Scannapieco\inst{1}\thanks{email to: cscannapieco@aip.de} \and Peter Creasey\inst{1} \and 
Sebasti\'an E. Nuza\inst{1} \and Gustavo Yepes\inst{2} \and Stefan Gottl\"ober\inst{1} \and Matthias Steinmetz\inst{1}}

\institute{Leibniz-Institut f\"ur Astrophysik Potsdam (AIP), An der Sternwarte 16, D-14482, Potsdam, Germany \and
Grupo de Astrof\'{\i}sica, Universidad Aut\'onoma de Madrid, Madrid E-28049, Spain}

\date{Received date /
Accepted date }

\abstract {}
{
We study the two main constituent galaxies of a constrained simulation 
of the Local Group as candidates for the Milky Way (MW) and Andromeda (M31).
We focus on the formation of the stellar discs and its relation to
the formation of the group as a rich system with two massive galaxies, 
and investigate the effects of mergers and accretion
as drivers of morphological transformations. We also assess
the effects of varying the assumed feedback model on our results by 
running two different simulations, a first one where only supernova feedback
is included and a second  where we additionally model radiation pressure from
stars.
 }  {We use a state-of-the-art hydrodynamical code which includes
star formation, feedback and chemical enrichment to carry out our study. 
We
use our two simulations, where we include or neglect the effects of
radiation pressure from stars, to investigate the impact of this process
on the morphologies and star formation rates of the simulated galaxies. } {
We find that the simulated M31 and MW have different formation histories, even though both inhabit, at $z=0$, the same environment.
These differences directly translate  into and explain variations in their
star formation rates, in-situ fractions and final morphologies. 
The simulated M31 candidate has an active merger history, as a result of 
which its stellar disc
is unable to survive unaffected until the present time.
In contrast, the  MW candidate has a smoother history with
no major mergers at late times, 
and forms a disc that grows steadily; at $z=0$
the simulated MW has an extended, rotationally-supported disc which
is dominant over the bulge.
Our two  feedback implementations predict
similar evolution of the galaxies and their discs, although some variations
are detected, the most important of which is the formation time of the discs: 
in the model with weaker/stronger feedback the discs form earlier/later. 
In summary, by comparing the formation histories of the two simulated galaxies,
we conclude
that the particular merger/accretion history of a galaxy 
rather than its environment at the LG-scales is the main
driver of the formation and subsequent growth or destruction of galaxy discs. } {}


\titlerunning{A constrained simulation of the Local Universe}
\authorrunning{Scannapieco et al.}

\maketitle

\section{Introduction}

The
formation of 
late-type spiral galaxies in the context of the  $\Lambda$-Cold Dark Matter (\lcdm)  cosmological framework is a 
long standing problem in astrophysics.
In 
hierarchical models such as \lcdm, the presence or absence of a strong, 
rotationally supported stellar component is influenced by processes acting at many scales as galaxies 
evolve;
baryons condense in the centres of dark matter haloes whilst retaining a considerable 
fraction of their angular momentum set at large scales \citep{Fall_Efstathiou_1980} and
a 
combination of viscous and dissipative processes in the gaseous component 
allows the formation of a disk from which a fraction is converted into a stellar 
component. Thicker stellar components result from dynamical processes such as 
disc instabilities \citep{Sellwood02} and the close encounters and mergers 
\citep[e.g.][]{Quinn93} that are ubiquitous in hierarchical growth models such 
as \lcdm, though some simulations \citep[e.g.][]{Bird13} alternatively suggest the 
older stellar disks may have been born thick.

Numerical simulations are the best tool to study galaxy formation
within \lcdm, and to investigate the
relationship between the formation of stellar discs and the
hierarchical build-up of the haloes containing them 
\citep[][]{S09,Sales12}.
During recent years, important progress has been made in the
field, and
simulations currently appear to
be reaching the level of sophistication
required to study the details of the assembly of galaxies and their stellar discs.
In fact, while a first generation of models
inevitably suffered from the angular momentum
catastrophe \citep[e.g.,][]{Navarro91,Steinmetz99}, more recent
codes showed that various types of feedback
provide galaxies with  self-regulated
star formation activity, in turn allowing the
formation of young, rotationally-supported stellar discs \citep[][]{S12}. 
Feedback processes produce a decoupling between the 
formation of a  gas disc after its condensation
within a halo and
the onset of star formation activity; in this way, 
stellar 
discs can form  later, in more quiescent periods rather than during
the early, violent times characteristic of the formation
of the haloes.
When  discs form later, they also have a greater 
chance to survive and can grow to form
extended, highly-rotating systems by $z=0$ \citep[][]{S09}.
At the heart of the problem of disc formation/survival is therefore
the problem of understanding 
how star formation proceeds in galaxies, and what
are the  possible regulating mechanisms and their
relative efficiencies.

For Milky Way-mass galaxies, the most efficient
feedback mechanism is believed to be that produced
by stars, via the injection of
energy into the interstellar
medium during different phases of stellar evolution.
In particular, various  models
for feedback from supernovae (SNe) 
have been included in simulations in
the last decade, and demonstrated that SN feedback
is indeed capable of regulating star formation
and of producing
stellar discs with high angular momenta 
\citep[e.g.,][]{Okamoto05,Governato07,S08,S09,Piontek11}. 
However, stellar discs in these models are usually subdominant
over the
massive bulges that form early. 
The more recent inclusion of additional feedback 
in the form of
radiation pressure  produced by massive stars showed that this process
provides significant kinetic energy to the gas, particularly at early
times, helping to produce even younger, more massive galaxy
discs
similar to those observed \citep[][]{Aumer13,Stinson13,Hopkins14,Agertz15}.

Despite the recent progress, 
the Aquila code comparison project \citep{S12} showed
that, even when the formation history of a halo is fully specified,
different models predict the formation of a galaxy with different
gas fractions, morphologies, stellar masses and angular momentum content.
Aquila clearly demonstrated
that the primary uncertainty in these types of simulations is related to
the processes of the unresolved baryonic physics that result in star formation and 
feedback and their respective
modeling,
and generated consensus on the need to produce
similar results using different codes, despite
their individual successes.
Efforts to overcome these problems are indeed ongoing,
both in terms of direct comparison between codes \citep{AGORA}
and of the development of more physically-motivated routines
to treat star formation and feedback
(e.g., \citealt{Christensen12,Agertz13,Aumer13,Creasey13,Stinson13,Vogelsberger13,Agertz15,Hopkins14,Murante15,Trujillo15}).

An important motivation to study the formation of late-type spiral galaxies
is the membership of our own galaxy, the Milky Way,
within that family. 
The MW 
provides an
ideal place to constrain galaxy formation theories,
as we have  detailed observational data 
on a large number of  stars residing in its different stellar components.
A  salient characteristic of the MW is
that it lives in a group environment,
known as the ``Local Group" (LG), with
another  large spiral, Andromeda (M31), 
lying at less than a Mpc distance from it.
A number of observational studies showed that the immediate environment
of galaxies
indeed plays a role in the determination of their properties
at the scales of galaxy clusters and rich groups
\citep[e.g.,][]{Dressler80, Hermit96, Guzzo97,Blanton03,Girardi_2003},
although for systems like the LG this is
more controversial \citep[e.g.,][]{Bahe_2013,Ziparo_2013}.
On the other hand, simulation studies 
have almost exclusively used initial conditions (ICs)
that follow the formation of a galaxy which is
isolated from massive neighbours at the Mpc scales \citep[see, however,][]{Few12}.
A study on possible effects of environment on the formation
of the MW and M31 is therefore required in order to understand
if these are representative
of galaxies of a similar mass and to better interpret
observational results.

In this work, we use constrained ICs of the Local Universe,
where a LG-like pair of galaxies form. We study the formation and evolution
of these two LG-galaxies, candidates
for the MW and M31, using a state-of-the-art hydrodynamical
code for cosmological simulations. 
We use these simulations to investigate 
how the LG galaxies could have formed, and to look for 
possible imprints of the formation of the LG on their final
properties. In particular, we focus
on the formation of the main stellar components
of the galaxies, and investigate possible correlations
between the formation of stellar discs 
and large-scale effects such as merger events and misalignment between the
gaseous and stellar discs. 
The main questions we want to address here are whether 
our simulation predicts the formation of two large
spiral galaxies similar to the MW and M31 and
what are the mechanisms determining/affecting their formation. 
We will show that only the simulated MW has a stable stellar
disc 
that survives to $z=0$; while in M31 various mergers
continuously affect the disc, and it produces
a galaxy with a much lower level of rotational support.
We find similar results using two different versions
of the code that we use to test whether results
are affected by the particular
treatment of feedback processes.
It is worth noting that, as shown in the Aquila Project \citep{S12}, 
the final
morphology of a galaxy depends sensitively  
on the amount of star formation during the early epochs of
the formation of galaxies, and therefore on the treatment of feedback
processes which regulate the star formation activity.

Other studies using our simulations are
presented in a series of companion papers.
\cite{Nuza14} analysed
the properties of the neutral, warm and hot gas within and around
the haloes of the MW and M31 candidates in one of our simulations,
including accretion/ejection
rates of the different components and HI covering fractions
that we contrasted to observational results. 
In \cite{Creasey15}, we further investigate the relation
between environment and galaxy properties, by comparing the
properties of our simulated MW and M31 with those of
galaxies of similar mass but formed in more isolated environments.
Our findings suggest that some of the properties
of galaxies, such as the star formation rates, may
indeed be directly affected by the environment at the Mpc scales;
while others, like the disc stellar mass fraction, are more sensitive
to the merger and approach of satellite systems to the inner regions
of the galaxies. 
(For other simulation studies analysing the relation between 
galaxy properties and environment in constrained simulations 
-- albeit at much larger scales -- 
see e.g. \citealt{Nuza10,Nuza14b}).

The outline of our paper is as follows: in Section~\ref{sec:sims}
we describe the initial conditions and code used for our study;
Section~\ref{sec:LG} describes the assembly of the LG and
the LG galaxies in our simulations; and in Section~\ref{sec:mergers}
we discuss the morphological transformations of the simulated
galaxies during their evolution, and investigate the 
relation between the evolution of morphologies and the merger histories.
Finally, in Section~\ref{sec:conc}, we summarise our results.

\section{The simulations}
\label{sec:sims}

We use for this study a simulation designed to produce
a LG-like system in the right cosmological environment,
and investigate the formation of its two main constituent galaxies,
candidates for the Milky Way and Andromeda.
In this section we describe the numerical methods needed to carry out
such a study, 
and  summarize the numerical set-up of the simulations.

\subsection{Simulation code}

We use the Tree-PM Smoothed Particle Hydrodynamics (SPH) 
code {\sc gadget3} \citep{Springel05,Springel08}, and run
two different simulations where additional updates are included.
The first simulation, labelled ``CS'' throughout this paper,
uses the extensions of \cite{S05} and \cite{S06} for
star formation, metal-dependent cooling, chemical enrichment
and (thermal) feedback from Type II and Type Ia supernova (SN) explosions, 
a multiphase model for the gas component, and a UV background
field \citep{HM96}.
The second simulation, labelled ``MA'', additionally includes
the effects of radiation pressure from stars and kinetic
feedback from SN explosions, as described in \cite{Aumer13}.

We have shown in previous work that
the CS model is able to reproduce the formation 
of galaxy discs from cosmological initial conditions,
alleviating the angular momentum problem. However,
although discs have realistic sizes and angular momentum content, 
overly massive bulges are also formed which are usually dominant 
(in mass) over the discs (\citealt{S08, S09, S10, S11, S12}). 
The more recent MA implementation is an update
to the CS model. In this case, early
stellar feedback from new born stars and kinetic feedback
from SN explosions is additionally considered, which leads to stronger
effects particularly at high redshifts.
The MA model produces galaxies
with  a high angular momentum content, moderate stellar masses
and dominant discs, 
in better agreement with observations of spiral galaxies.
It is therefore relevant for our work to compare
the results obtained with the two models, which
allows us to test the robustness of our results, particularly
in relation to the effects of mergers  during
the evolution of galaxies.

In both models,
star formation takes place in dense ($n_{\rm H}> 0.03$ cm$^{-3}$)
regions of converging flow, 
with a star formation rate per unit 
volume equal to
\begin{equation}
\dot\rho_\star = c_*\,\frac{\rho}{\tau_{\rm dyn}},
\end{equation}
where  $\rho$ and  $\tau_{\rm dyn} = 1 / \sqrt{4\pi G\rho}$ are respectively the
density and dynamical time of the gas particle, and $c_{*}$
a star formation efficiency (that we set to $c_*=0.1$).
Our simulations use the multiphase model described in \cite{S06},
where  the neighbour list of any particle $i$ with entropy $A_i$
is constructed ignoring neighbours $j$ with $A_j<50A_i$ 
(provided they are not part of a shock to avoid unphysical
behaviour). In this way,
a multiphase medium is generated, where particles of varying
densities and temperatures (i.e. {\it cold} and {\it hot}) 
can coexist in the same spatial region. These cold and hot
phases are however not fixed, as the selection of particles
to be ignored as neighbours depends on the relative
properties of each pair of particles $i$, $j$ and is not
based on any pre-fixed values for different phases. 
As explained in \cite{S06}, not only
this model allows coexistence between a dense and a diffuse phase,
but also a more efficient deposition of the supernova energy.

In the CS model, each star particle can
``explode'' a maximum of two times, first as SN Type II (SNII)
and later on  as SN Type Ia (SNIa). During each SN explosion,
the star distributes chemical elements (according to the
corresponding SNII and SNIa chemical yields) and energy into
neighbouring gas particles. The energy and chemical production
of each exploding star is distributed in equal proportions
to  its hot (non star-forming) and cold
(star-forming) gas neighbours. We assume a  
canonical $0.7\times 10^{51}$ ergs of energy to be released per SN. 
The distribution of chemical elements occurs at the time of
the explosion, as well as the release of energy to the hot phase.
However, for the cold phase, feedback energy is accumulated in a reservoir,
until it is sufficient to allow the particle thermalize with
the local hot phase.
In practice, this works as follows: for each cold
gas particle, we calculate the average density and entropy over
its hot neighbours 
(those particles that were ignored
in its neighbour list according to our multiphase model)
and calculate the energy needed for the cold
particle to reach these values, $\Delta E$. This energy is then compared
to the energy stored in the reservoir $E_{\rm res}$, and when 
 $E_{\rm res}>\Delta E$ the cold particle gets the reservoir energy as
thermal energy.
Note that the value of $\Delta E$ is different and time-dependent for each cold
particle, 
which makes the code suitable to adapt to the  cold/hot regions with different
temperatures and densities, which are present at different times
during the formation of a galaxy and at different spatial regions.

The MA model works similarly to the CS model; however, in this case, the 
SN
energy is split into a kinetic and a thermal
component. 
The kinetic energy transferred from the exploding star to the gas neighbours
is calculated assuming 
that the momentum
of the SN ejecta is conserved, and characterized by an outflow
velocity $v_{\rm out}$:  $\Delta p=\Delta m\cdot v_{\rm out}$.
We use $v_{\rm out }=3000$ km s$^{-1}$, with momentum given in the radial
direction. As the momentum is distributed into neighbouring particles,
which
typically translates into a 
velocity kick per particle of
$\Delta v_{\rm i}\sim 20-30$ km s$^{-1}$.
The remaining energy (i.e. the total energy released per
exploding star minus the kinetic energy that has been already
been transferred to the neighbouring gas particles)
is stored into the  thermal reservoir of the
gas receiving feedback, and is treated in the same way as in the CS model.
The MA model includes a treatment for radiation pressure, where
 the rate of momentum deposition in the gas is parametrized 
as
\begin{equation}
\dot{p_{\rm rp}} = (1+\tau_{\rm IR}) {L(t)\over{c}}
\end{equation}
with $L(t)$
being the UV-luminosity of the stellar population and $\tau_{\rm IR}$ an infrared
optical depth which depends on the density, metallicity and velocity
dispersion of the gas around the star. This last dependency makes the effects
of radiation pressure larger in systems with higher
velocity dispersions, typical of high-$z$ galaxies,
allowing a more efficient regulation of star formation
at early epochs and higher star formation levels
at late times.

Note that, as described in \cite{Aumer13}, the MA model
considers additional updates 
(new treatment of chemical enrichment and metal diffusion,
AGB stars, and element-by-element cooling)
that we do not include here, 
and therefore the only differences between our two simulations
is the inclusion of radiation pressure and the additional kinetic feedback
assumed for SNe.
In order to
better compare results from our two simulations,
we have also used the exact same input parameters for star formation and 
feedback when appropriate (which in some cases differ from those assumed
in \citealt{Aumer13}),
and also used the same choice for the gravitational softening, which
was (for the high resolution particles) $0.7$ kpc,  
fixed in physical coordinates  since $z=3$ and fixed in 
comoving coordinates
at earlier times.
In this way, 
any difference in the results can be directly associated
to the effects due to the radiation pressure of young stars and 
kinetic feedback during the SN phase.

We note that some properties of the simulated galaxies, particularly
the detailed final morphology, are expected to be sensitive
to   resolution \citep{S11} and
 to 
the choice of the gravitational softening \citep{Aumer13}. 
However, although higher resolution or/and lower values for
the gravitational softening will allow the vertical/radial structure of discs
to be better resolved, the presence or absence of a disc will not
be affected by these choices (as long as a minimum resolution
is considered). Our current work focuses on the evolution of
discs and therefore our results are still valid, even when 
the spatial resolution is not high enough to resolve the radial/vertical
structure of discs with a large number of resolution elements.

\subsection{Initial conditions and simulation set-up}

The initial conditions (ICs) used for this study are
constrained initial conditions  of the  Local Universe,
that are part of the CLUES (Constrained Local UniversE 
Simulations\footnote{http://www.clues-project.org/}) project.
The ICs  reproduce, by construction,  the known dynamical properties
of our local environment at the present-day 
(see \citealt{Gottloeber10} and \citealt{Yepes13} 
for details)
and
 use the zoom-in technique, 
in a periodic box of
$64$ $h^{-1}$Mpc   on a side. 
In the centre of this cube, a high-resolution region of 2$~h^{-1}$Mpc 
comoving radius contains dark matter and gas particles,
with a mass resolution of $2.8\times 10^6$~M$_\odot$  and  $5.6\times 10^5$~M$_\odot$ 
respectively. Outside the high-resolution region, 
the mass distribution is described by low-resolution
collisionless particles, as usual in zoomed simulations.

The ICs are consistent with 
a \lcdm\ 
universe with  WMAP-5 cosmological parameters:
  $\Omega_{\rm M}=0.279$ (matter density), 
 $\Omega_{\rm \Lambda}=0.721$ (dark energy density),
 $\Omega_{\rm bar} = 0.046$ (baryon density), $H_0=100 h$ km s$^{-1}$ Mpc$^{-1}$ with
$h=0.7$ (Hubble parameter), 
and $\sigma_8=0.8$ (normalization of the power spectrum),
and correspond to a  starting redshift of $z=50$.

\subsection{Milky Way and Adromeda candidates}

At $z=0$, the simulated LG has two main constituent
galaxies, with total masses  similar to the MW, 
that we take as candidates for the MW and Andromeda\footnote{
Note that, as the constraints can only be imposed on large scales, 
the properties of the two main galaxies
will not necessarily resemble those of the actual MW and M31 in detail.}. 
We will refer to these galaxies as G1 (candidate for M31) and G2 
(candidate for MW)\footnote{Note that in \cite{Nuza14} we instead refer 
to G1 and G2 as M31$^{\rm c}$ and MW$^{\rm c}$ respectively.}
and, where appropriate, we will add an acronym CS/MA to denote
the simulation run.
The assignment of our galaxy candidates to the real counterparts is based 
on their mass, G1 being the most massive at $z=0$ (see Table~\ref{tab:prop}). 
Additionally, with this choice, we recover a similar relative orientation 
between the galaxies in comparison to observations \citep{Nuza14}.

Table~\ref{tab:prop} shows the  main $z=0$ properties
of G1 and G2 in our two runs:  $R_{200}$,
defined as the radius where the density contrast is 200 times the
critical density, and the total ($M_{200}$), gaseous ($M_{\rm gas}$) and stellar 
($M_{\rm stars}$) masses, all within $R_{200}$.
The present-day  masses of 
G1 ($\sim$$1.7\times 10^{12}\,\,$M$_\odot$) 
and G2 ($\sim$$1.2\times 10^{12}\,\,$M$_\odot$)
are similar to the values estimated
for Andromeda ($\sim$$1.4\times 10^{12}\,$M$_\odot$; \citealt{Corbelli10,Watkins10})
and  the Milky Way ($\sim$$10^{12}\,$M$_\odot$;  
\citealt{Wilkinson99,Smith07,Watkins10,Bovy12,Kafle12,Piffl14}) from observational studies.

\begin{figure*}
  \begin{center}

{\includegraphics[width=4.2cm]{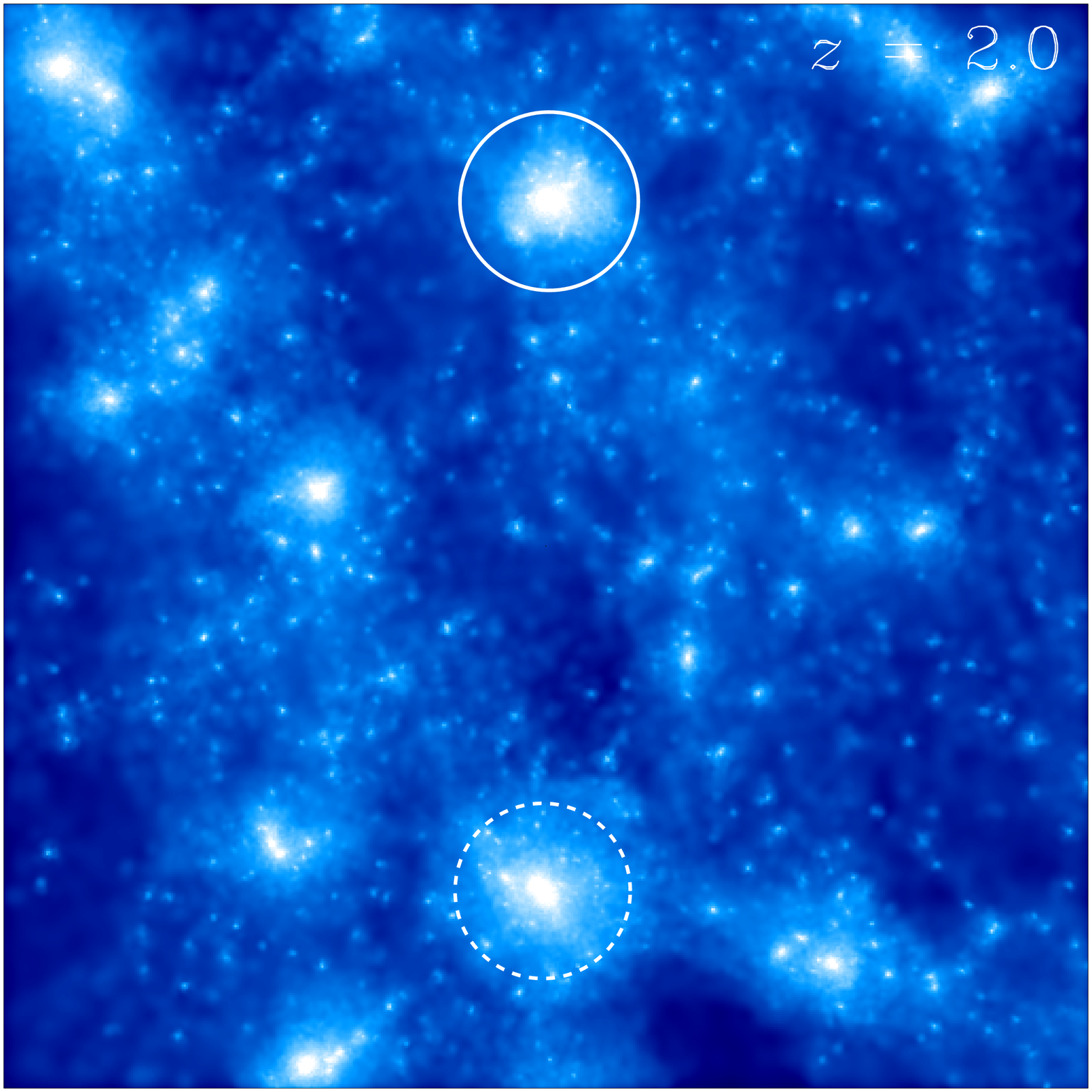}\includegraphics[width=4.2cm]{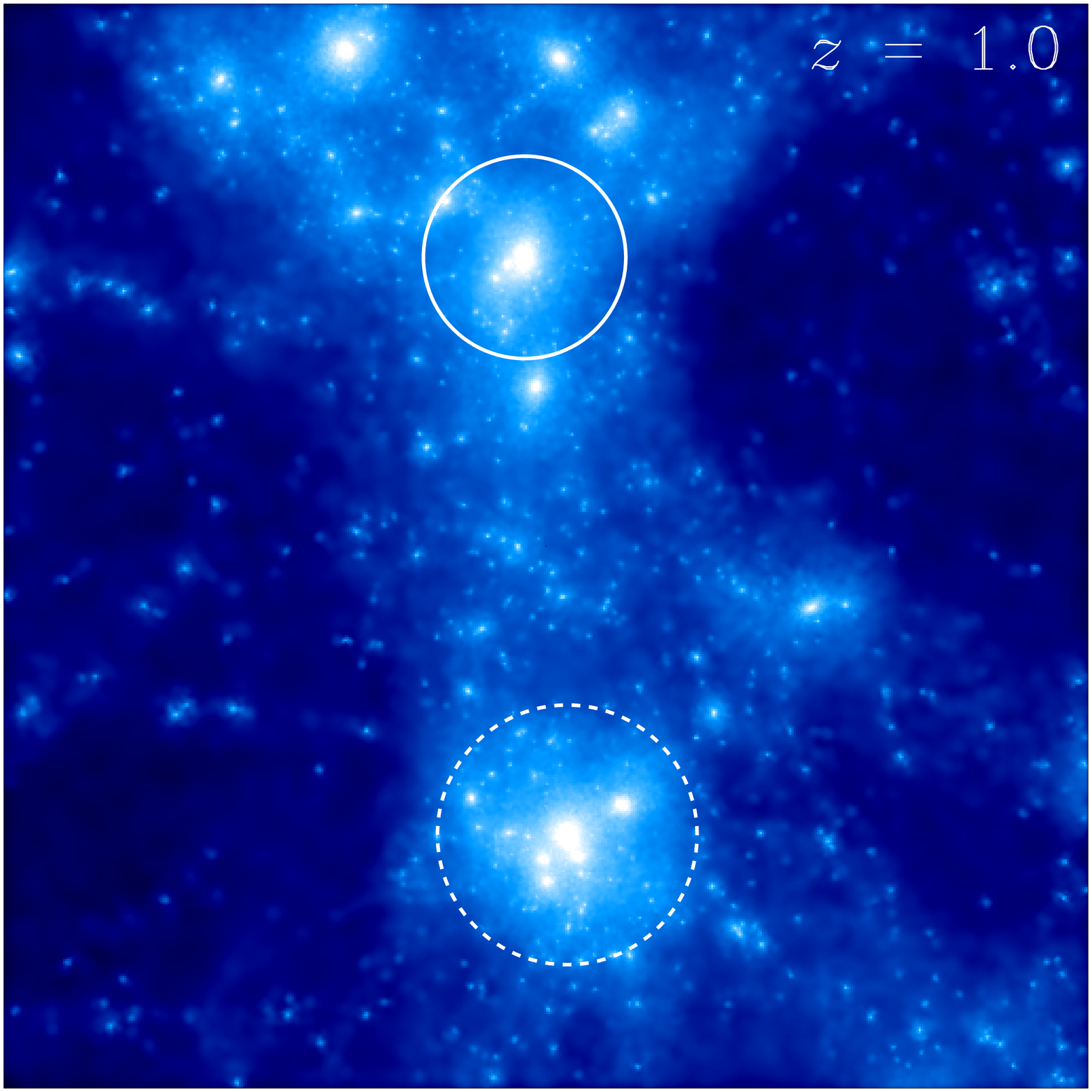}\includegraphics[width=4.2cm]{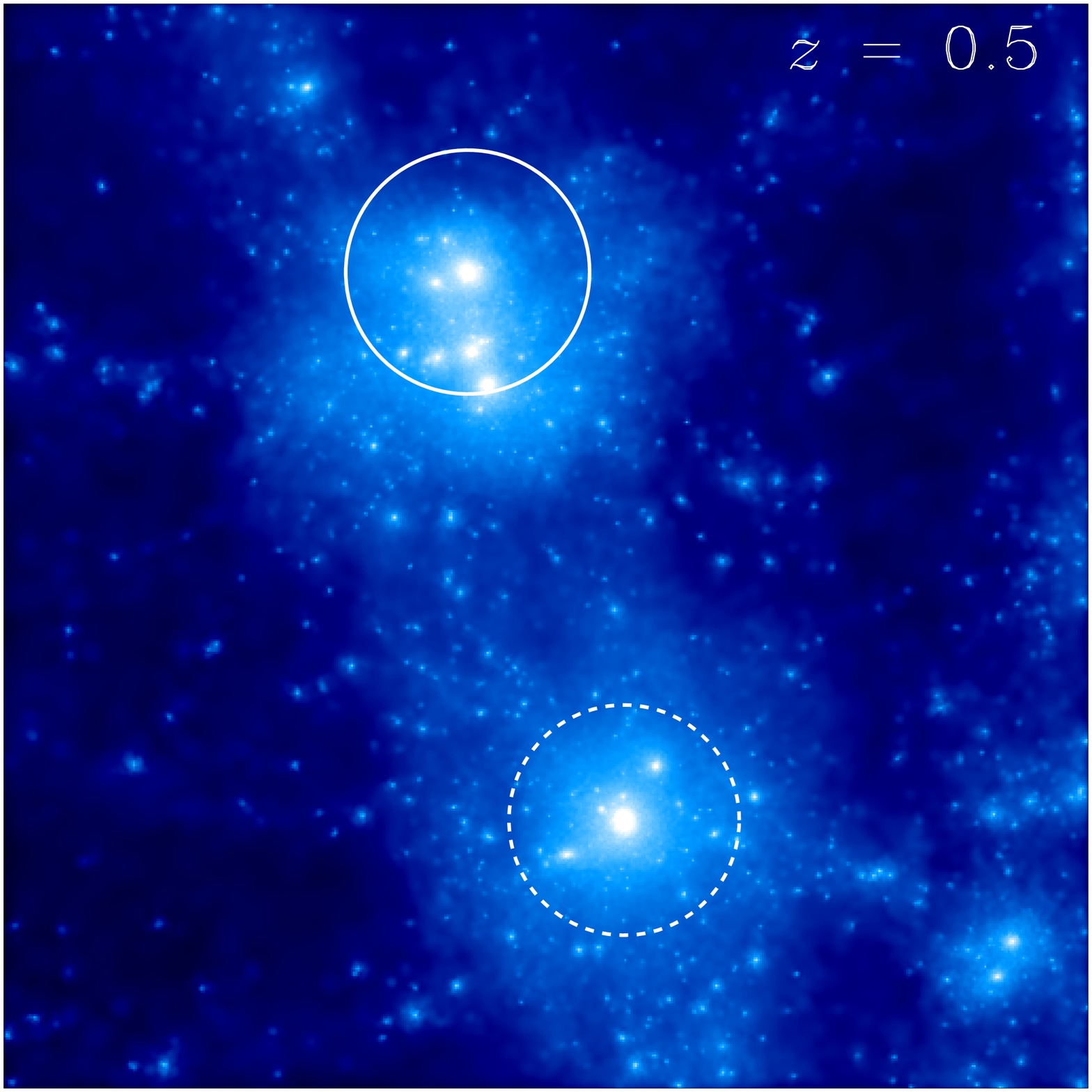}\includegraphics[width=4.2cm]{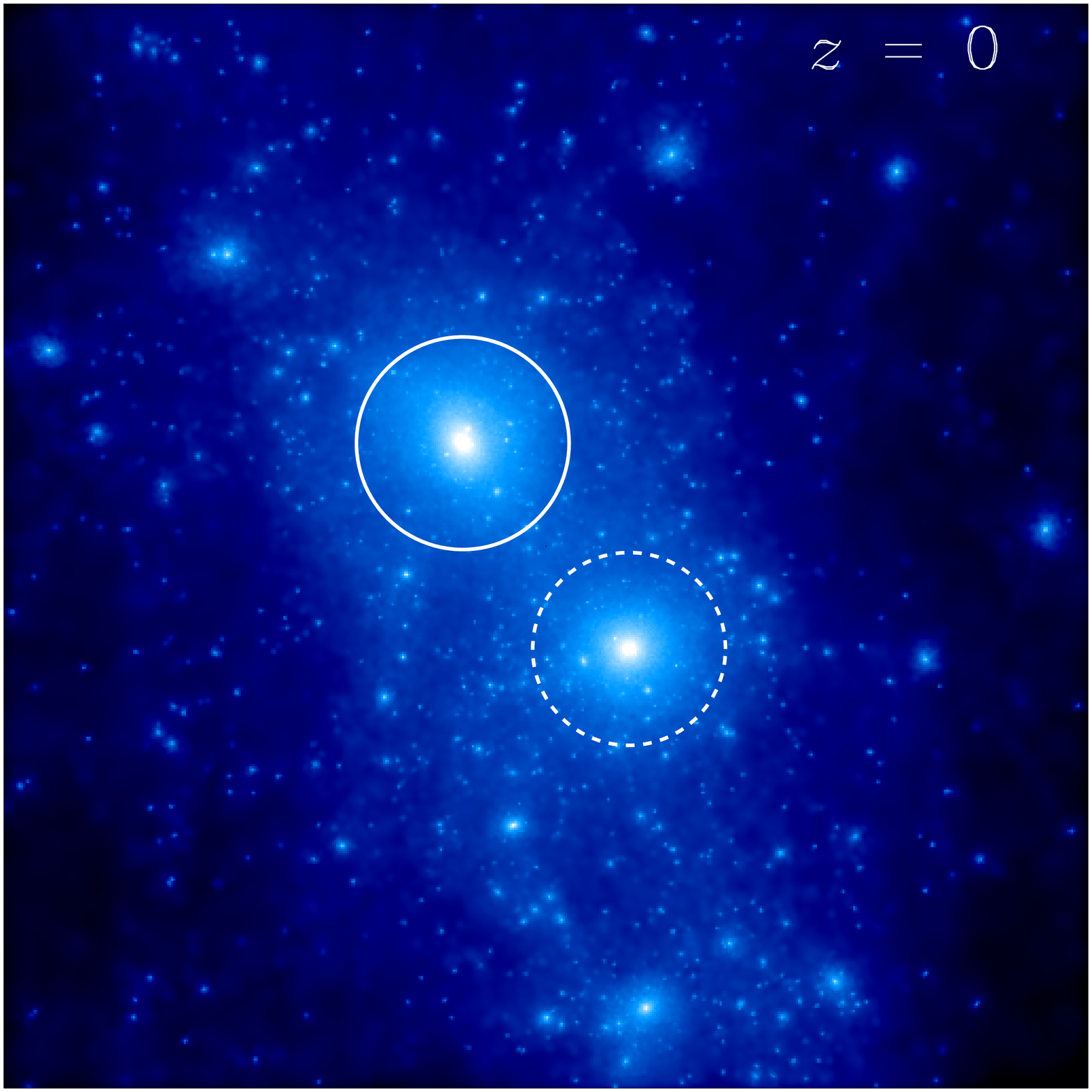}\includegraphics[height=4.2cm]{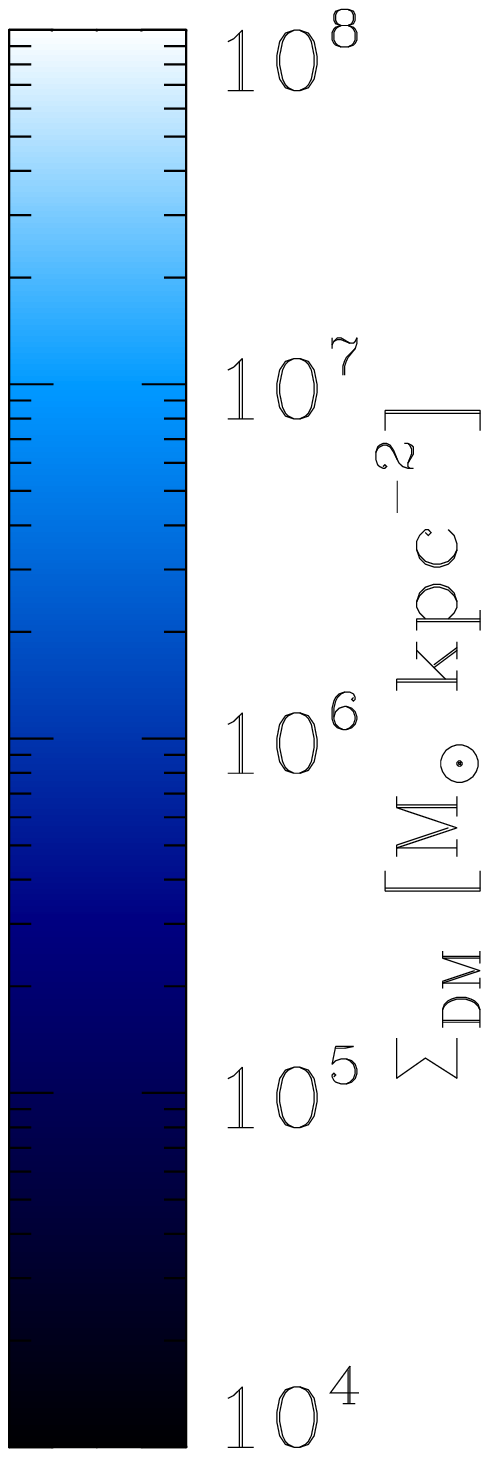}}

{\includegraphics[width=4.2cm]{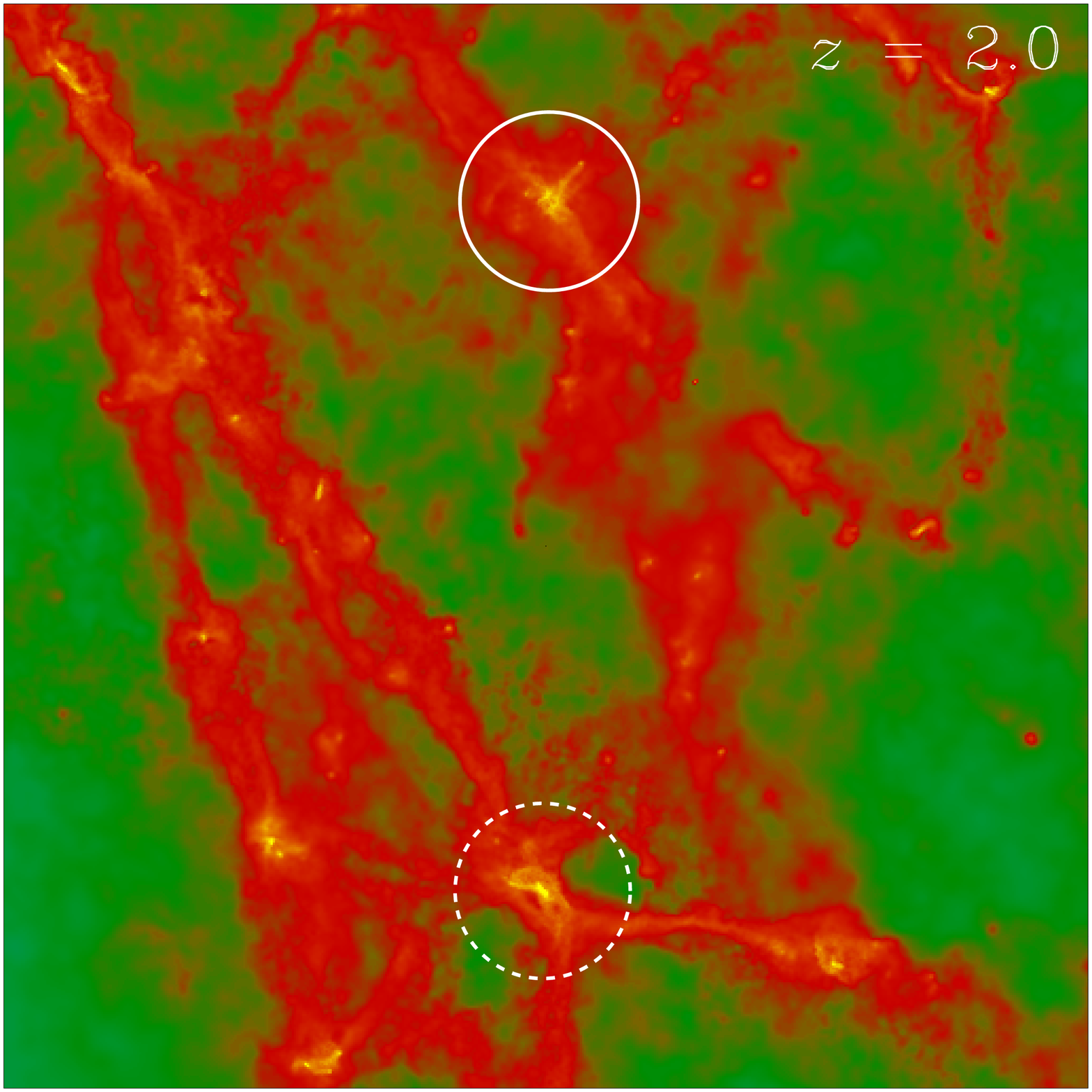}\includegraphics[width=4.2cm]{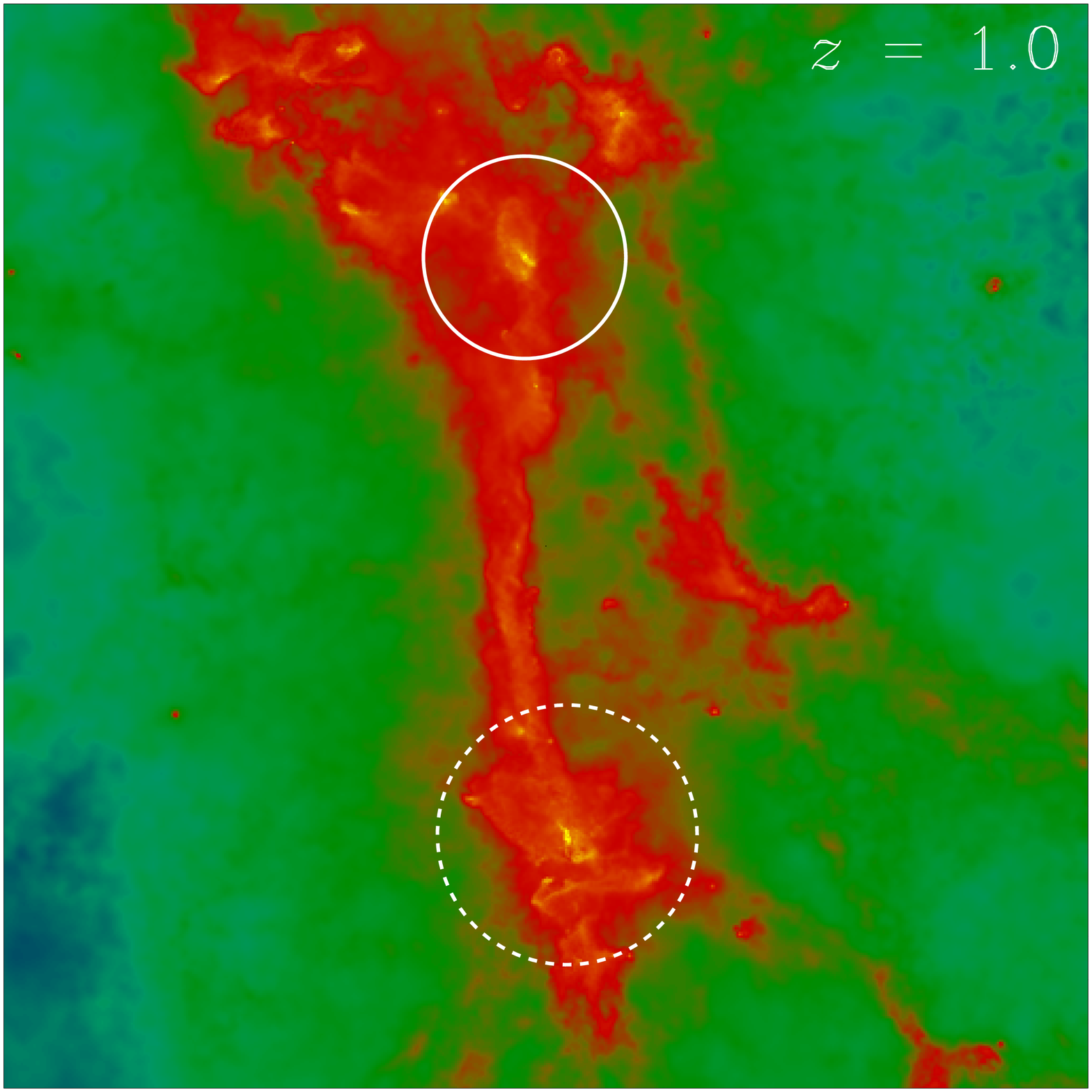}\includegraphics[width=4.2cm]{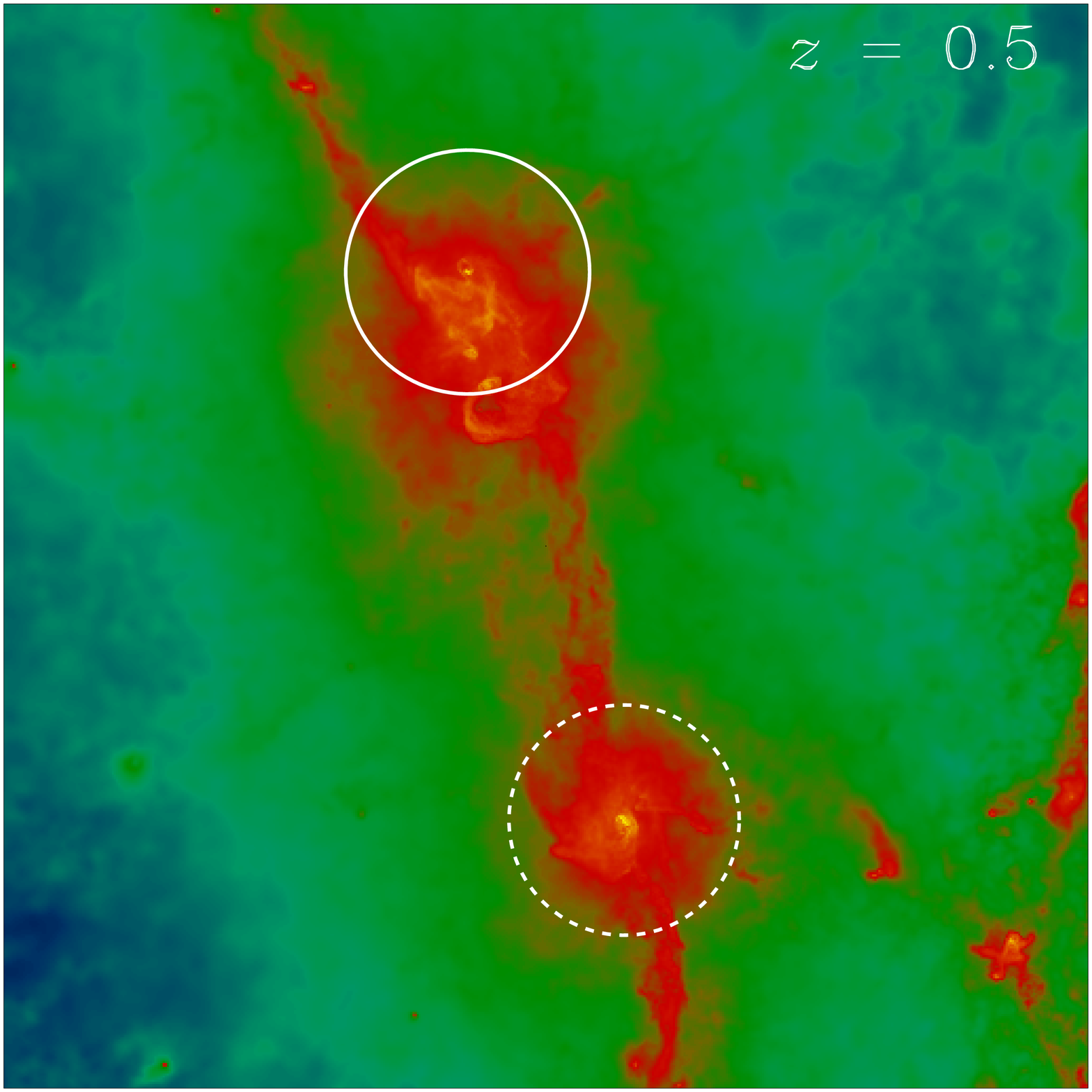}\includegraphics[width=4.2cm]{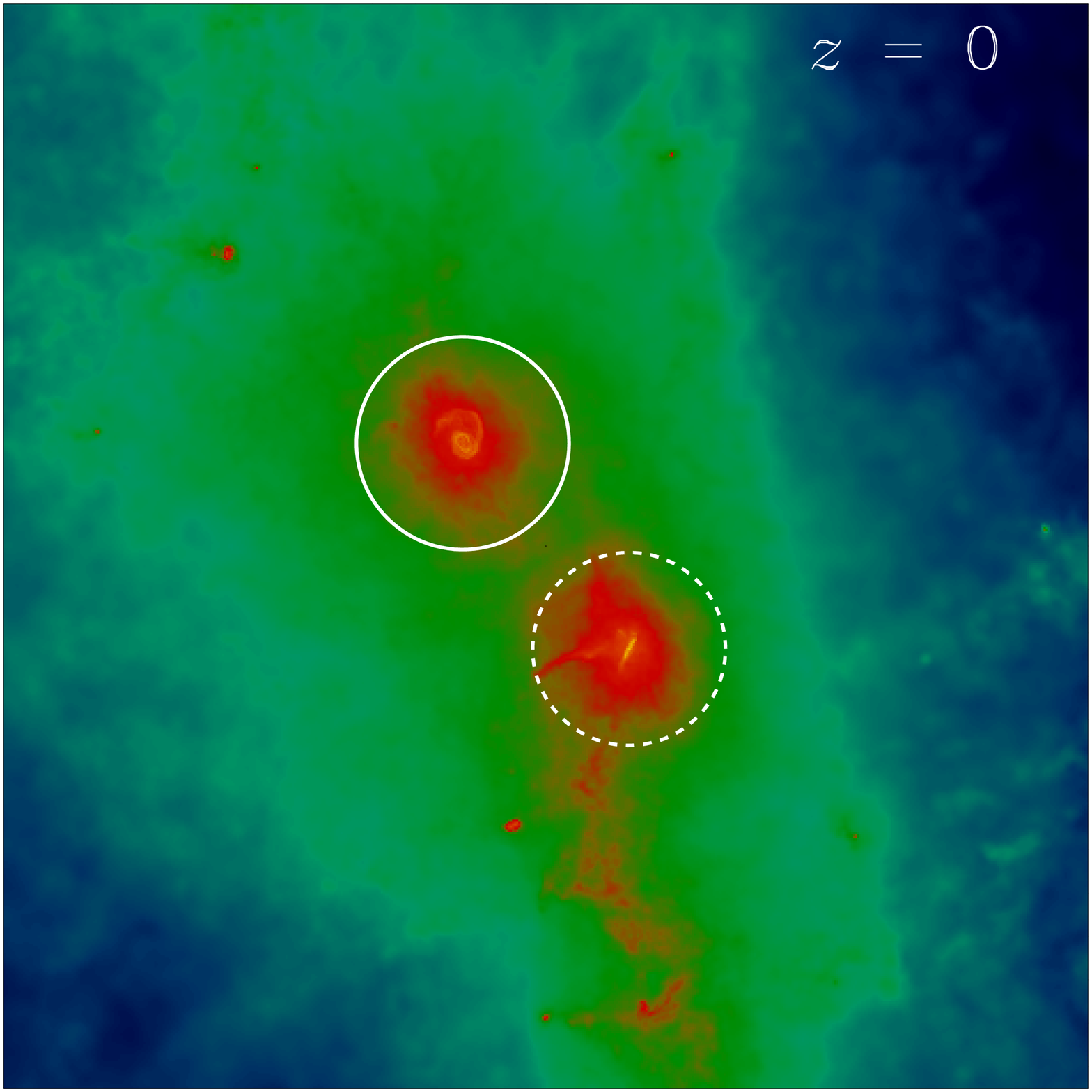}\includegraphics[height=4.2cm]{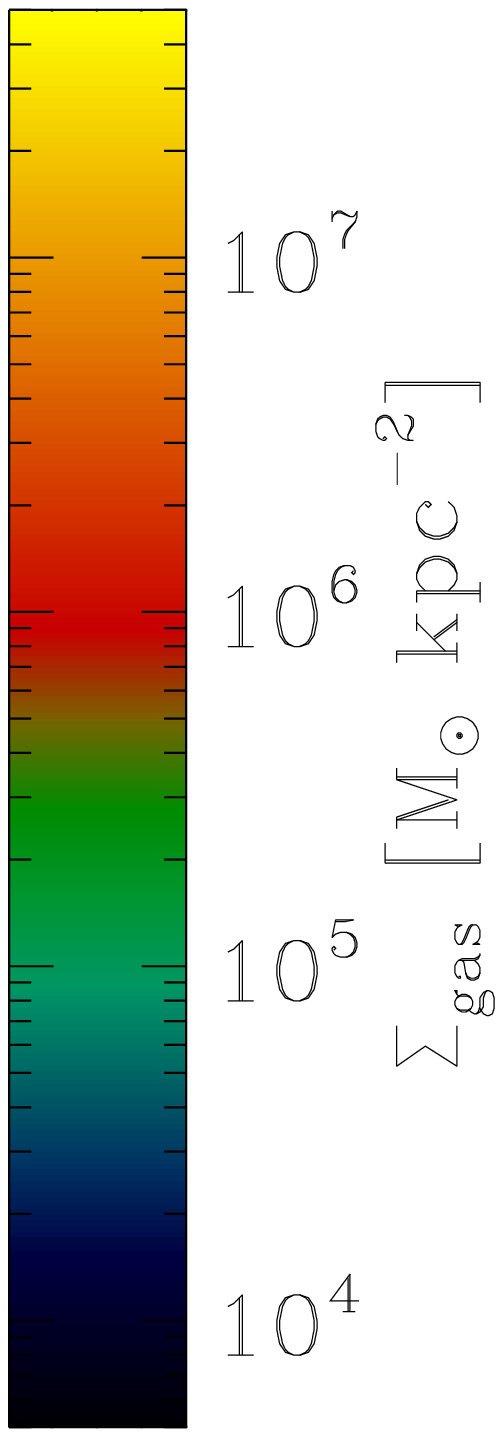}}

{\includegraphics[width=4.2cm]{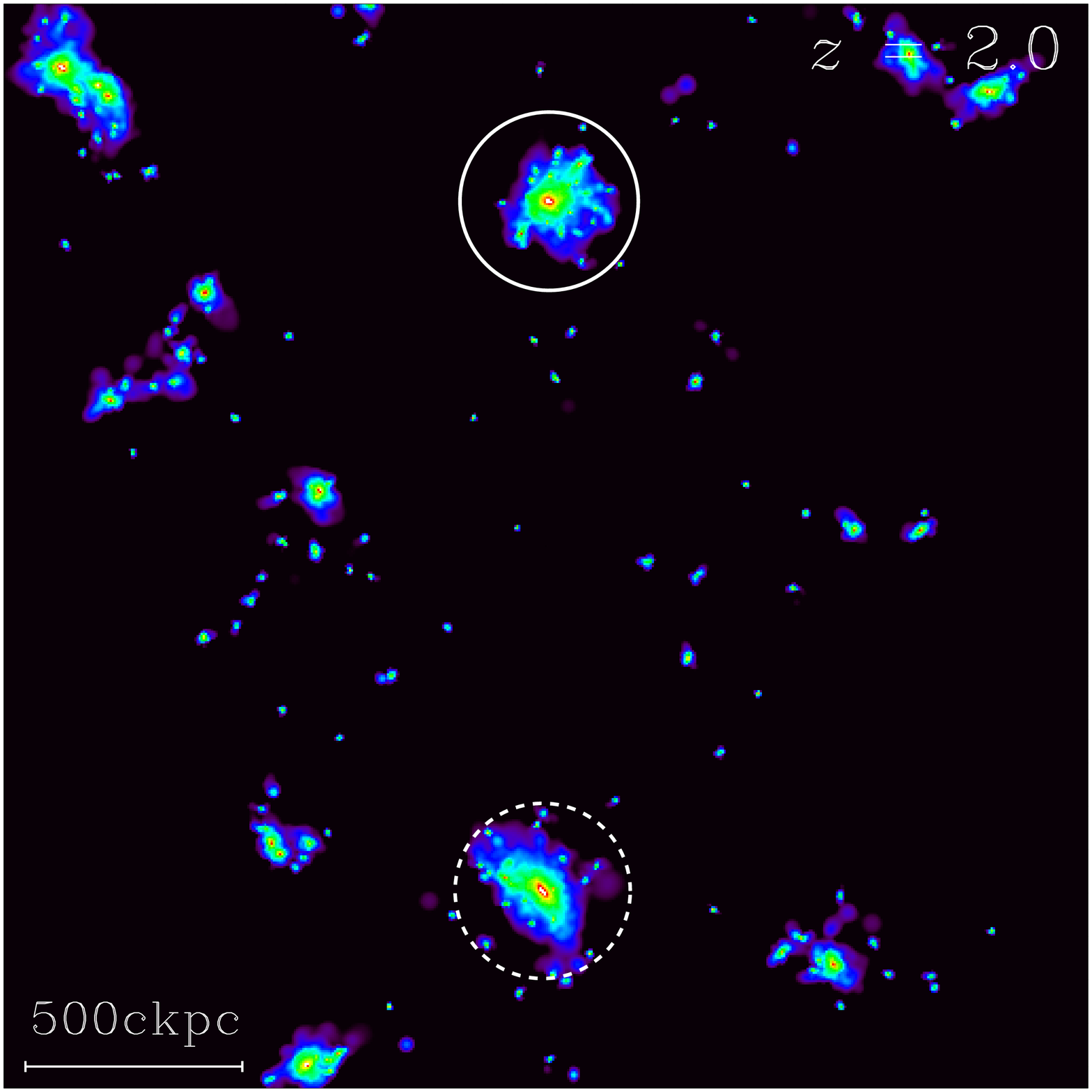}\includegraphics[width=4.2cm]{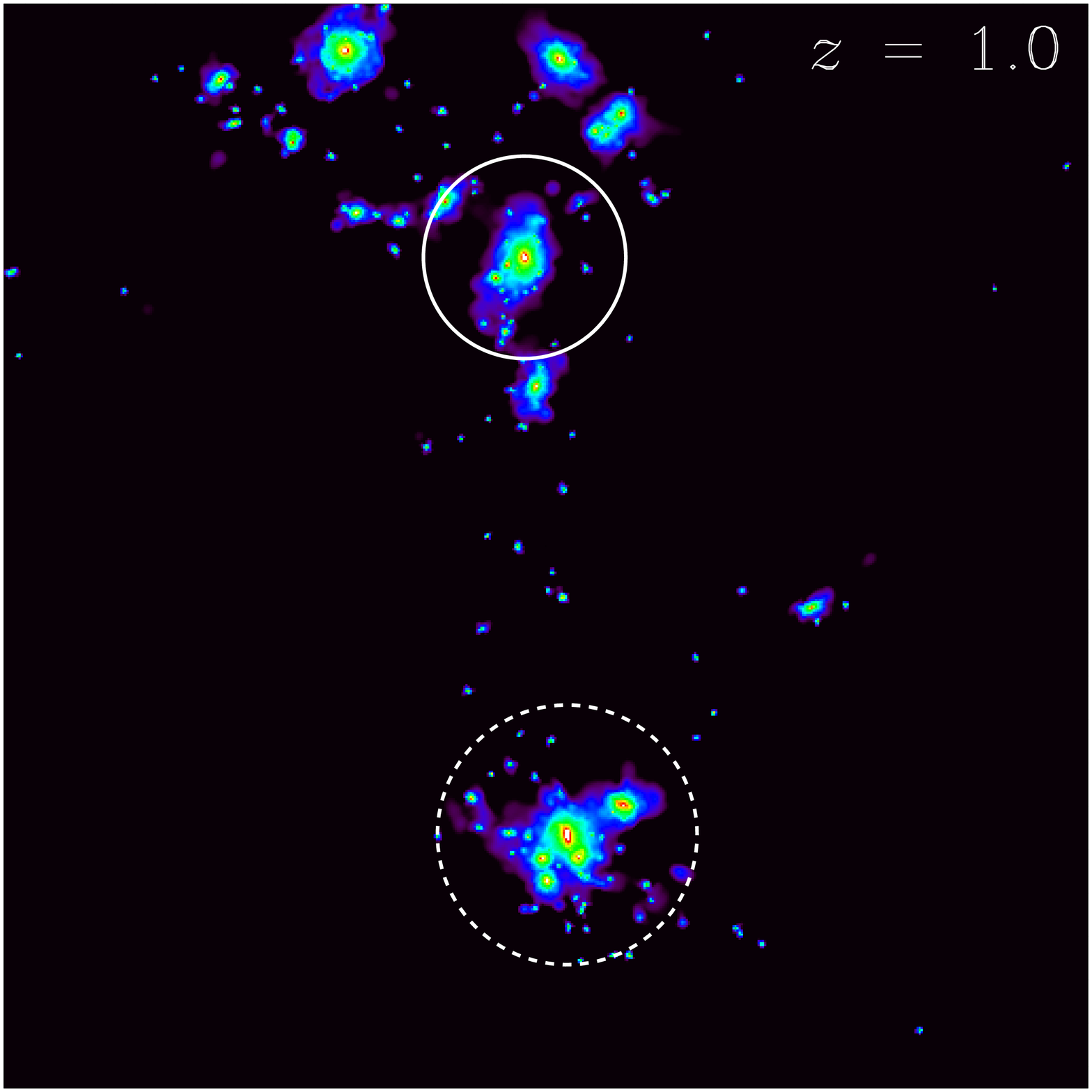}\includegraphics[width=4.2cm]{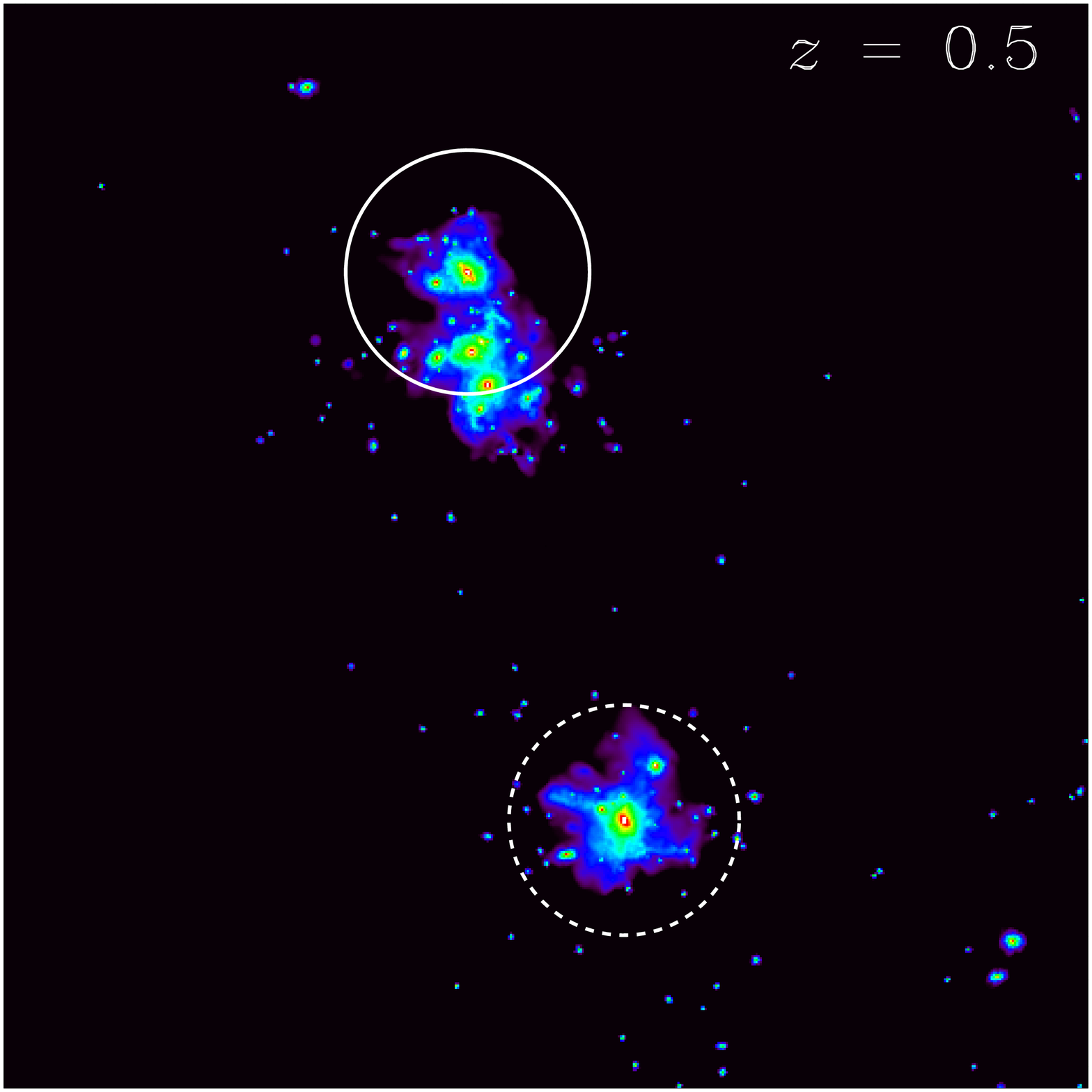}\includegraphics[width=4.2cm]{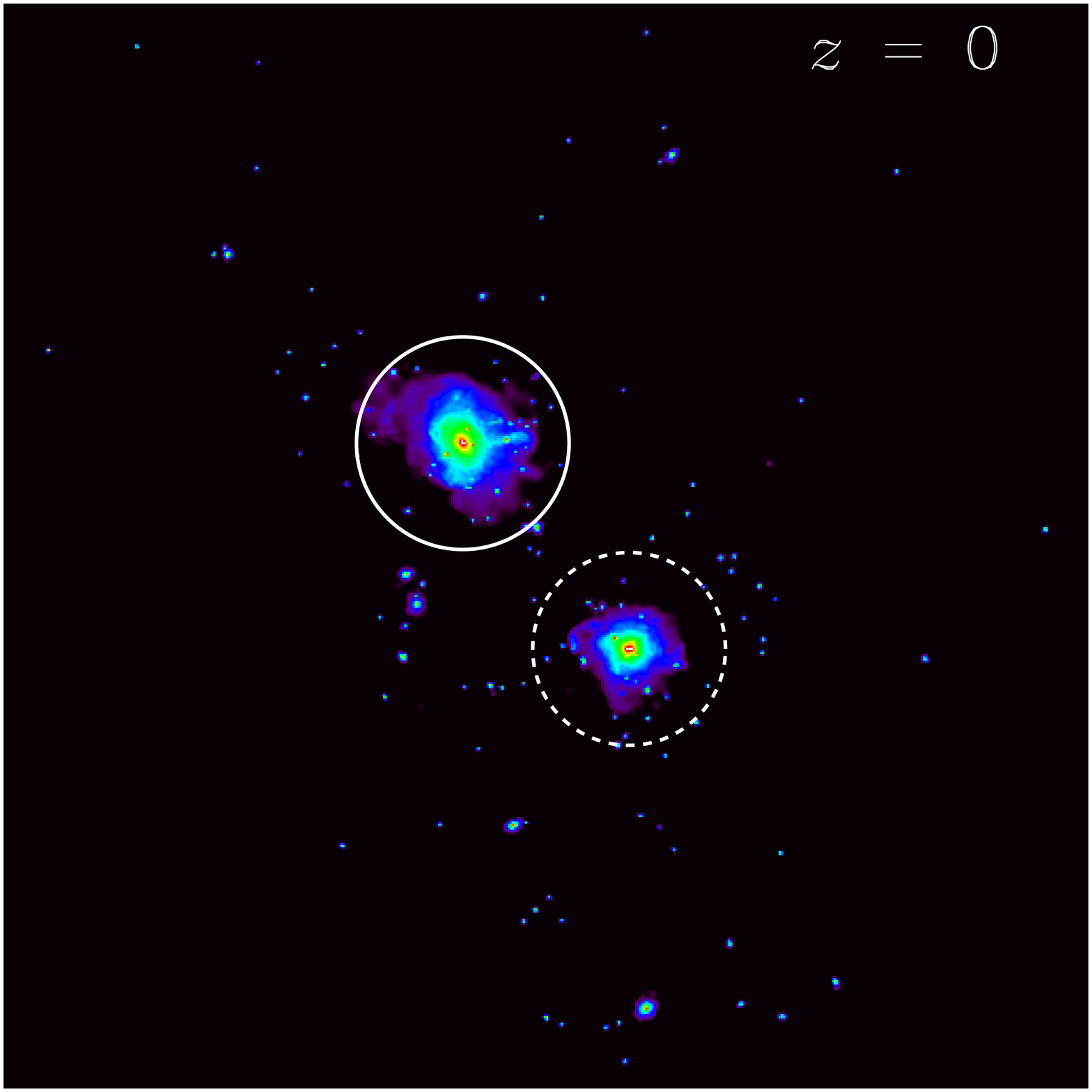}\includegraphics[height=4.2cm]{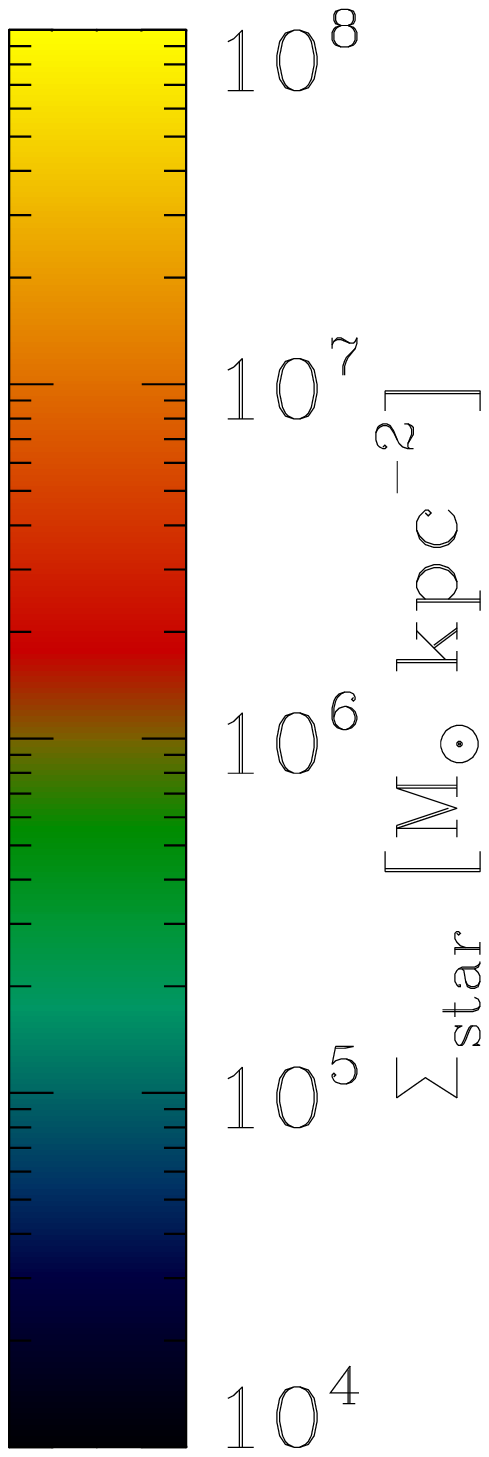}}

\caption{Spatial distribution (in an arbitrary projection) 
of the dark matter (upper panels), gas (middle panels) and stars (lower panels) 
in our simulation, in a cube of 
 2.5 cMpc on a side and for redshifts $z= 2,  1,  0.5$ and $0$. The $R_{200}$  of G1 (solid line) and G2 (dashed line) at each redshift are indicated, and the colour scale (of each mass component) is fixed in order to highlight differences in the distributions at different times. 
The plots correspond to run CS, but results at these scales are very similar for both simulations. }
\label{fig:maps_large_scale}
\end{center}
\end{figure*}

\section{The assembly of the LG and the LG galaxies}\label{sec:LG}

The assembly history of the LG and of the
LG galaxies (G1, candidate for M31 and G2, candidate for MW)
is seen in 
Fig.~\ref{fig:maps_large_scale}, where we show the
dark matter, gaseous and stellar  mass
distributions  (in an arbitrary projection)
 at $z=2,  1, 0.5$ and $ 0$. 
The different panels of Fig.~\ref{fig:maps_large_scale} are centred
at the centre of mass  of the G1-G2 system at the corresponding
cosmic epochs, and the two  galaxies are surrounded
by solid (G1) and dashed (G2)  circles  indicating their $R_{200}$ radii. 
At each redshift, the plots  show the mass distributions in a cubic box
of  $2.5$ cMpc\footnote{For short, throughout the paper we will use a prefix {\it c} 
to length units to denote the use of comoving coordinates.} 
side length that corresponds to the high-resolution
region. The different colours span 4 orders of magnitude in logarithmic scale,
and we adopted the same colour scale at all redshifts in order to highlight
time variations in the distributions\footnote{These figures correspond to
 our CS run although, at these scales, the results are
qualitatively the same as those obtained with the MA model.}.

\begin{table} 
\begin{small}
\caption{Main properties of the G1 and G2 galaxies in our two simulations at $z=0$:
  radius $R_{200}$, and total, gaseous and stellar
 ($M_{200}$, 
$M_{\rm gas}$ and
   $M_{\rm stars}$) masses within within $R_{200}$.  }
\label{tab:prop}
\begin{center}
\begin{tabular}{lccccccc}
\hline

Galaxy-   & $R_{\rm 200}$  & $M_{\rm 200}$ & 
 $M_{\rm gas}$&$M_{\rm stars}$ \\
Model       &  {[kpc]}& { [$10^{10}$ M$_\odot$]} & 
 { [$10^{10}$ M$_\odot$] }& 
{ [$10^{10}$ M$_\odot$]} \\ 
\hline

\\

G1-CS   &  244.9&       167.9 &       6.67&       7.95\\
G1-MA    &  239.1&       155.1 &       6.13&       6.76\\
\\

G2-CS    &  222.2 &       125.1 &       5.69&       6.24\\
G2-MA    &  219.5 &       120.6 &       6.39&       5.48\\

\hline

\end{tabular}
\end{center}
\end{small}
\end{table}

At $z=2$, the characteristic filamentary structure at Mpc scales
is clearly seen; G1 and G2 are $550$ kpc apart (i.e. more than a Mpc
apart in comoving coordinates), and belong to different filaments.
 The Local Group evolves quickly and, by $z=1$, defines a system
where the mass distributions of its two main galaxies
are interconnected --  a feature that is particularly visible
in the gas component --  and have
collapsed to the same filament. The two galaxies
continue separating from each other until
$z\sim 0.5$ when they reach a maximum
separation of $\sim 900$ kpc. After this time, 
G1 and G2 start approaching each other, to reach  a final separation
of about $800\,$kpc.
The stars are located within the very centres of the dark matter haloes,
resulting in much less extended distributions compared to the gaseous 
and dark matter counterparts. A significant number
of smaller galaxies and satellites of G1 and G2 are observed, particularly
at high redshift; most of them have been accreted by the G1 and G2 haloes
by $z=0$.

\begin{figure*}
\begin{center}
\includegraphics[width=17cm]{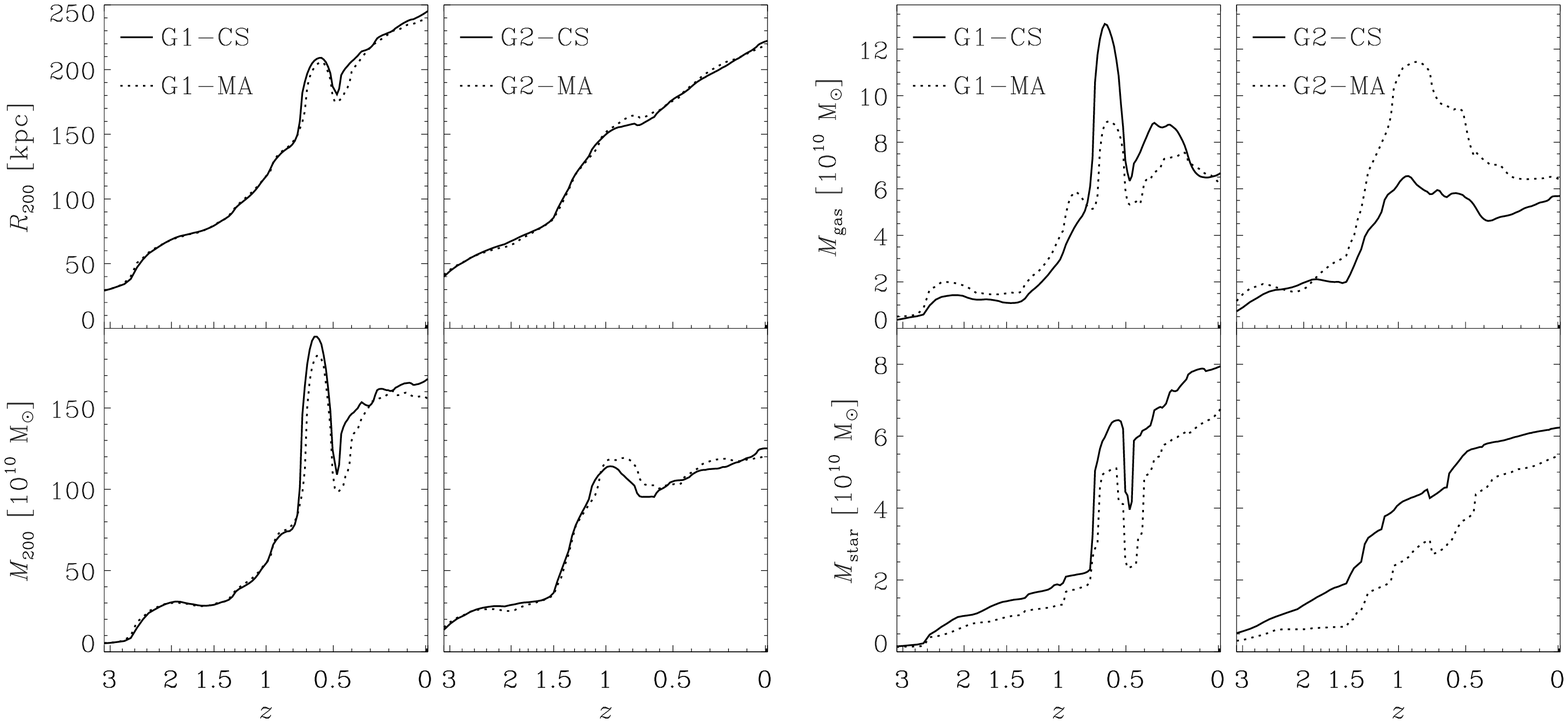}

\vspace{-0.5cm}
\caption{The size/mass evolution of G1 (solid line) and G2 (dashed  line) as a function of redshift. We show the evolution  of 
$R_{200}$ (upper-left panels), the total mass ($M_{200}$, lower-left panels), 
 the gas mass ($M_{\rm gas}$, upper-right panels) and the stellar mass ($M_{\rm star}$, lower-right panels),  all within $R_{200}$.
For G1, the peak seen in all panels at $z \approx 0.65$ is owing to the emergence of two large objects 
within $R_{200}$ (see further discussion in Section \ref{sec:mergers}).}

\label{mass_evol}
\end{center}
\end{figure*}

Although G1 and G2 are, at $z=0$, both part of the simulated LG, their
respective growth history are somewhat different, as 
evidenced from the evolution of their total masses and radii shown in  
the left-hand panels of Fig.~\ref{mass_evol}.
In the case of G1, the size and mass increase rapidly from $z=3$
to $z=0$, particularly
after $z\sim 1$ owing to the accretion of a number of satellite
galaxies\footnote{We discuss the merger histories of G1 and G2 in Section~\ref{sec:mergers}.}. These satellites are being accreted at approximately the same
time around $z=0.9-0.7$, and 
contribute  a significant amount
of mass to G1, determining the increase and variations of $R_{200}$.
Slightly after $z=0.5$, 
the satellite galaxies 
are in their last apocentre, 
and the evolution of
G1 becomes more quiet until the present time.
In contrast, G2 has a more steady evolution at all times, with the most
significant changes at $z\approx 1-1.5$, which also result from
the entrance of a relatively small satellite galaxy.
The agreement between the evolution of G1 and G2 in our
two simulations is excellent in terms of the 
masses and 
radii.

The different growth of the haloes of G1 and G2 also reflect
in the assembly of the baryonic components, with a much smoother
evolution of G2 compared to G1, as shown in the right-hand panels of
Fig.~\ref{mass_evol}.
However, as the evolution of the gaseous/stellar masses are affected
not only by the growth of the haloes with time, but also by the transformation
of gas into stars and the return of mass from the stars to the gas
owing to stellar evolution,
larger differences between the two runs are found for the baryonic
components\footnote{Note, however, that the simulations
presented here only consider mass return from SN explosions, while they do 
not include a treatment for the mass return 
of intermediate-mass stars that would result in a larger correction.}.
As a result of the stronger feedback assumed in model MA,
galaxies have in general
 lower stellar and higher gaseous masses
in this run, compared to simulation CS.
This effect is already noticeable at high redshift, due to 
the early
feedback associated to massive, short-lived stars, that acts shortly after the onset
of star formation activity in the galaxies.

\begin{figure*}
\begin{center}
\includegraphics[width=15cm]{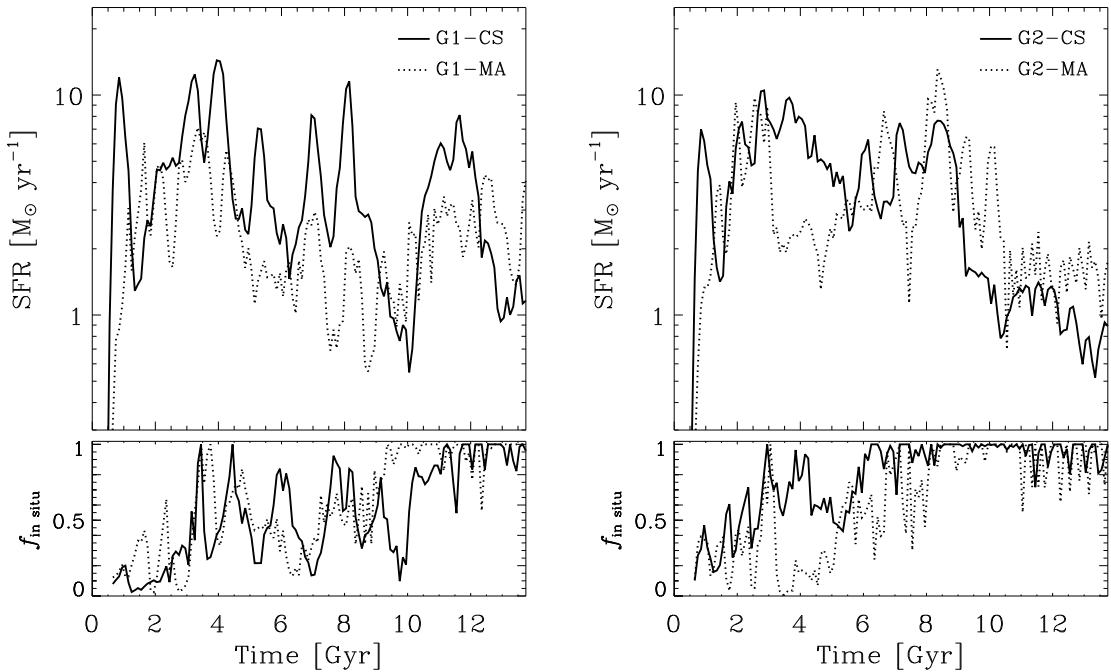}
\vspace{-0.5cm}
\caption{Star formation rate of G1 (left-hand panel) and G2 (right-hand panel) 
         as a function of cosmic time since the Big Bang for simulation runs CS (solid lines) and MA (dotted lines). The lower panels show the corresponding  in-situ fractions, i.e. the stellar
mass fraction formed in the main progenitor at each time interval.}
\label{fig:sfr}
\end{center}
\end{figure*}

The differences in stellar/gas masses  between results of runs CS and MA
are a consequence of the different 
star formation rates (SFRs), 
shown in the upper panels of Fig.~\ref{fig:sfr}.
To calculate the SFRs, we considered all stars that
end up in the main galaxy at $z=0$ (i.e. within 30 kpc),
and therefore include stars formed in-situ (within
the main progenitor at each time) and in satellite
systems that were later accreted.
The SFRs are higher at early times and gently
decline at more recent epochs.
Feedback effects start before in run MA compared
to run CS, as radiation pressure acts right after
the onset of star formation. In fact, the 
first star formation burst produced in run CS (in both
galaxies) is absent
in simulation MA.
In the case of  G1, we find that the
SFRs in model MA are in general lower compared to model CS,
with the most important differences at very early times
and around $6-8$ Gyr. For G2, differences are
significant (additionally to the very early epochs) at
$4-6$ Gyr, with lower SFRs for model MA,
and after $9$ Gyr
 of evolution, with  higher SFRs for model MA compared
to CS.

The lower panels of Fig.~\ref{fig:sfr} show the corresponding
in-situ fractions, $f_{\rm in-situ}$, which help us to understand
variations between the evolution of G1 and G2 that result from the different
SFRs of the progenitor galaxies and their satellites. 
An important difference between G1 and G2 
(in both runs) shows up very clearly in these plots: G1 has much
lower in-situ fractions at all times compared to
G2. Only at low redshifts does  G1 have $f_{\rm in-situ}\sim 1$,
while at earlier times there is a high contribution of stars
that form in  other systems than the main progenitor.
In contrast, G2 has higher in-situ fractions at all times:
$f_{\rm in-situ}\sim 1$ for at least the last $6-8$ Gyr.
As we discuss in  Section~\ref{sec:mergers}, these differences
can explain the origin of the final morphologies
of the simulated galaxies, as G1 has a much more active merger history
which makes its stellar disc much  more susceptible to partial or total destruction.

\section{Galaxy morphologies and their relation to the formation/merger history}\label{sec:mergers}

The figures of the previous section already revealed
that G1 and G2 exhibit differences in their growth/star formation history. 
In this section, we investigate how
these differences are
translated into variations in the final properties of the galaxies,
particularly in their morphologies/dynamical state.

\begin{figure*}
\begin{center}
{\includegraphics[width=3.2cm]{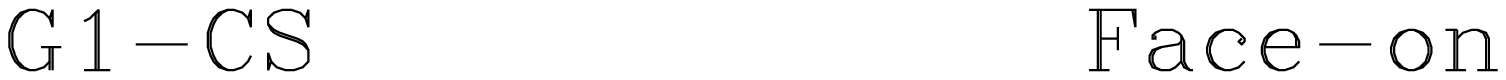}\includegraphics[width=3.2cm]{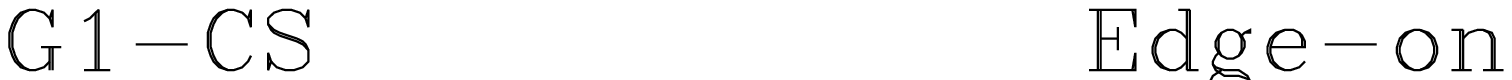}\hspace{1cm}\includegraphics[width=3.2cm]{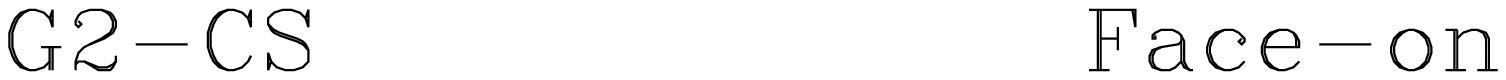}\includegraphics[width=3.2cm]{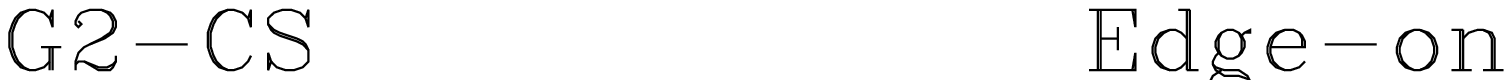}}
{\includegraphics[height=3.2cm]{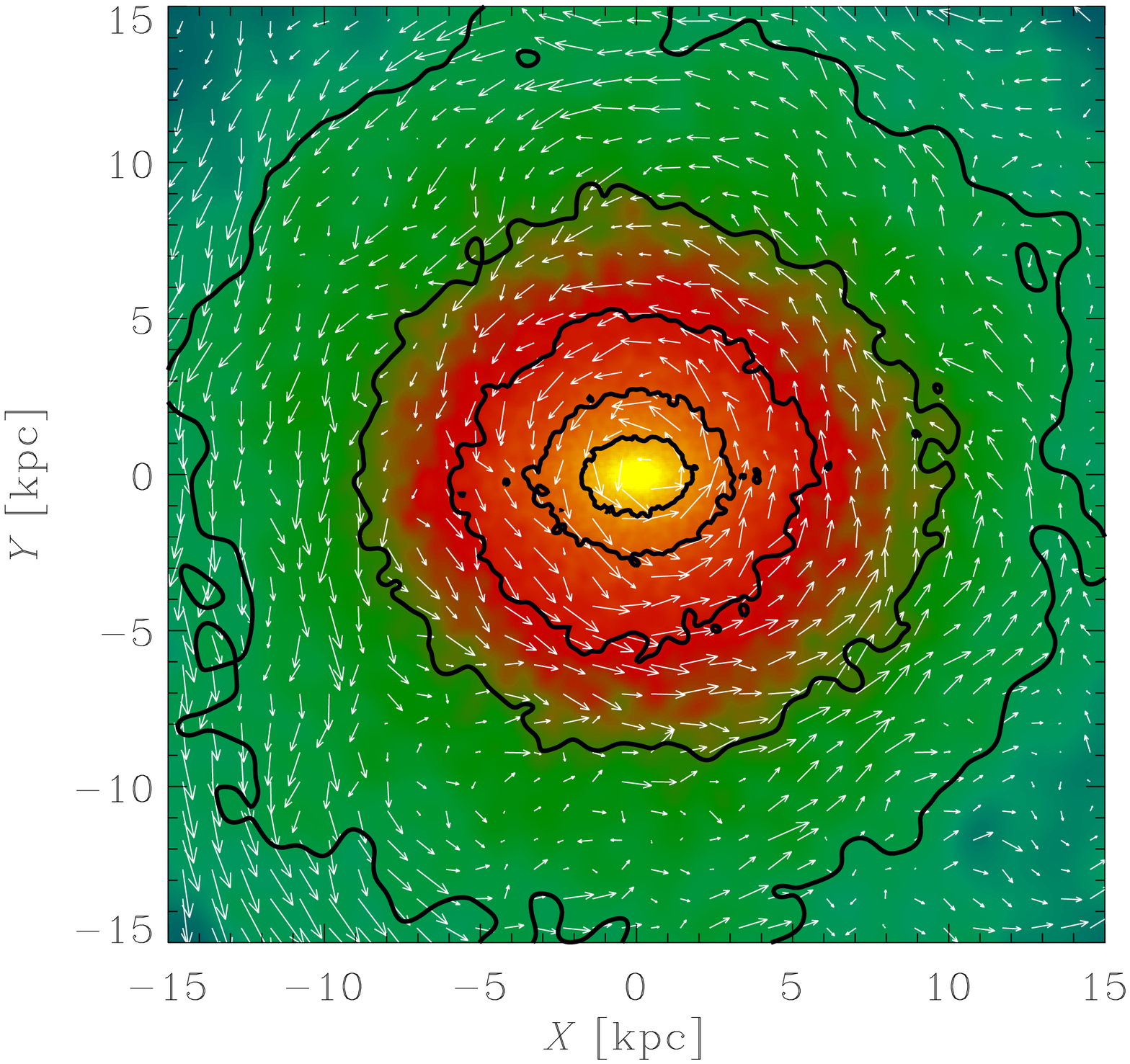}\includegraphics[height=3.2cm]{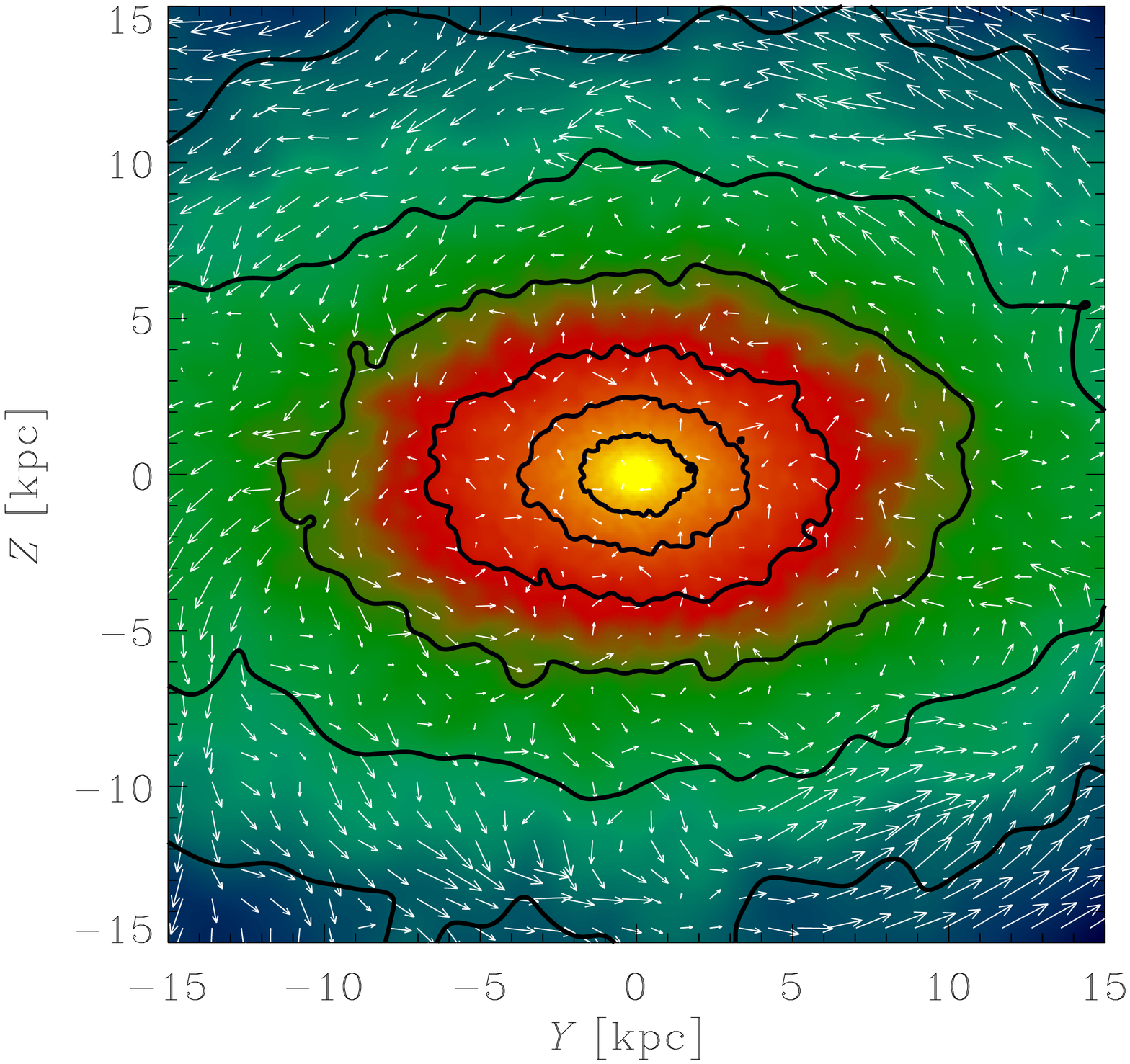}\hspace{1cm}\includegraphics[height=3.2cm]{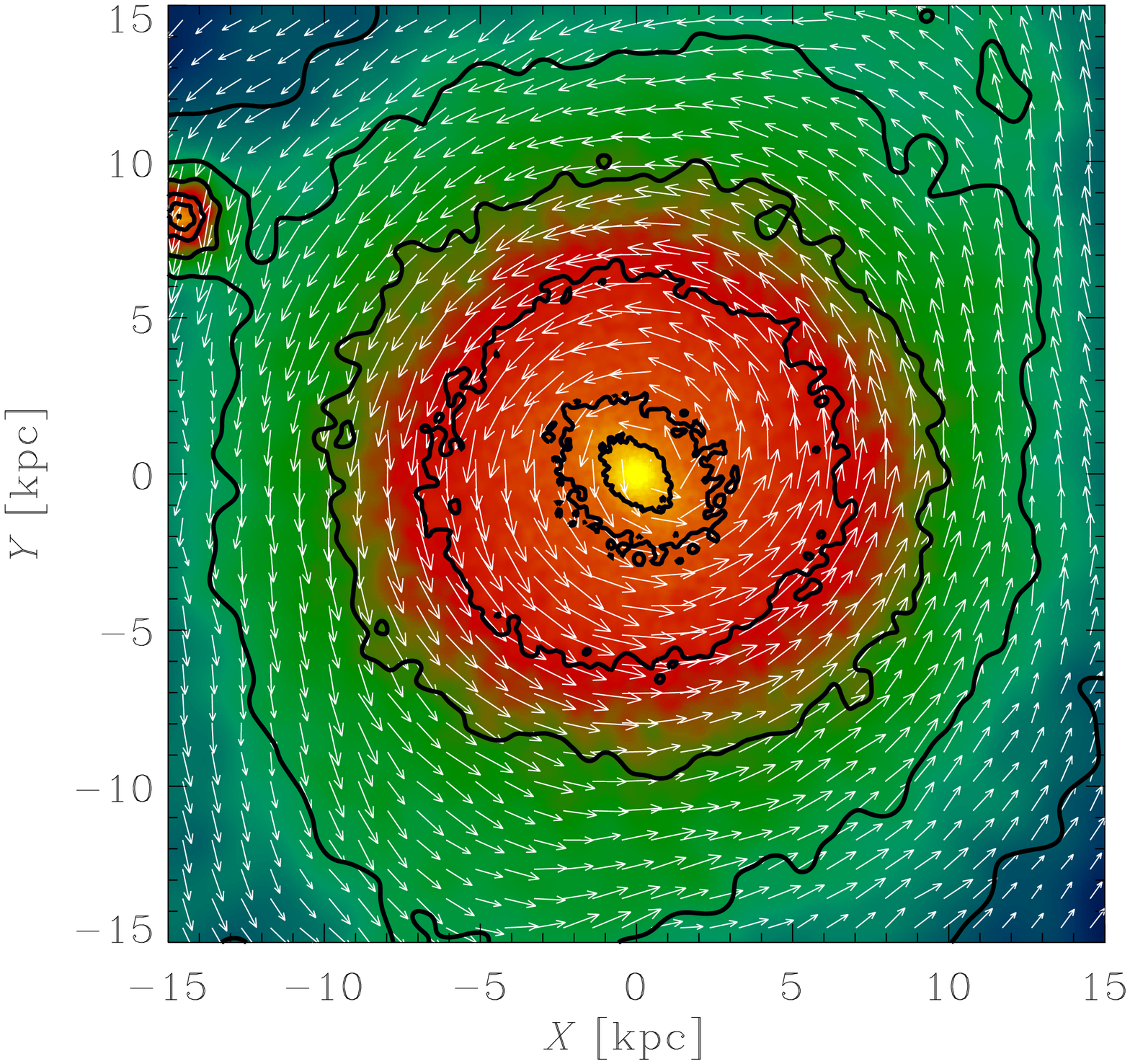}\includegraphics[height=3.2cm]{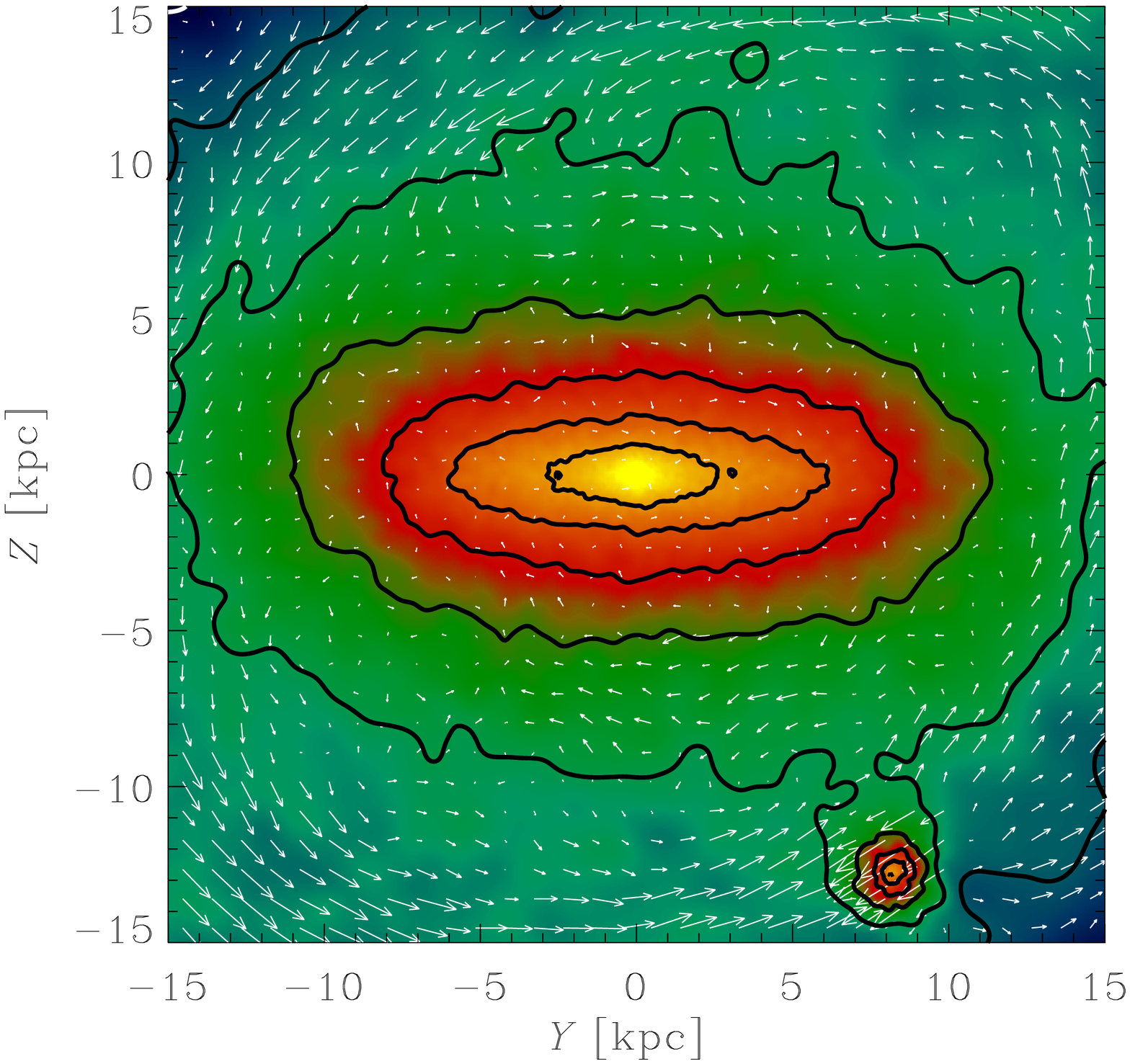}}

{\includegraphics[width=3.2cm]{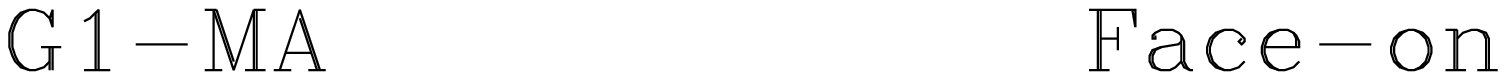}\includegraphics[width=3.2cm]{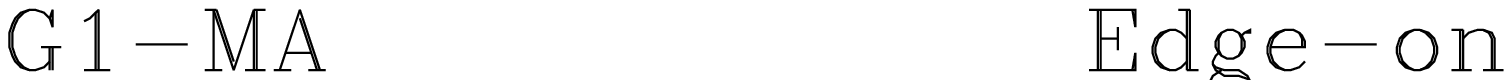}\hspace{1cm}\includegraphics[width=3.2cm]{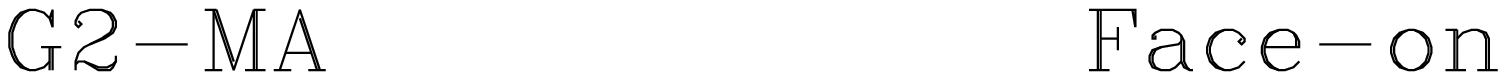}\includegraphics[width=3.2cm]{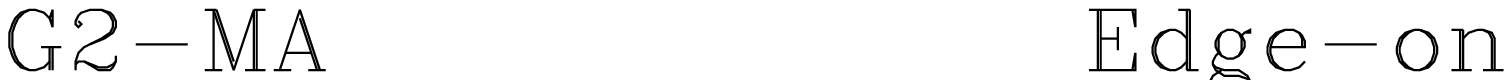}}
{\includegraphics[height=3.2cm]{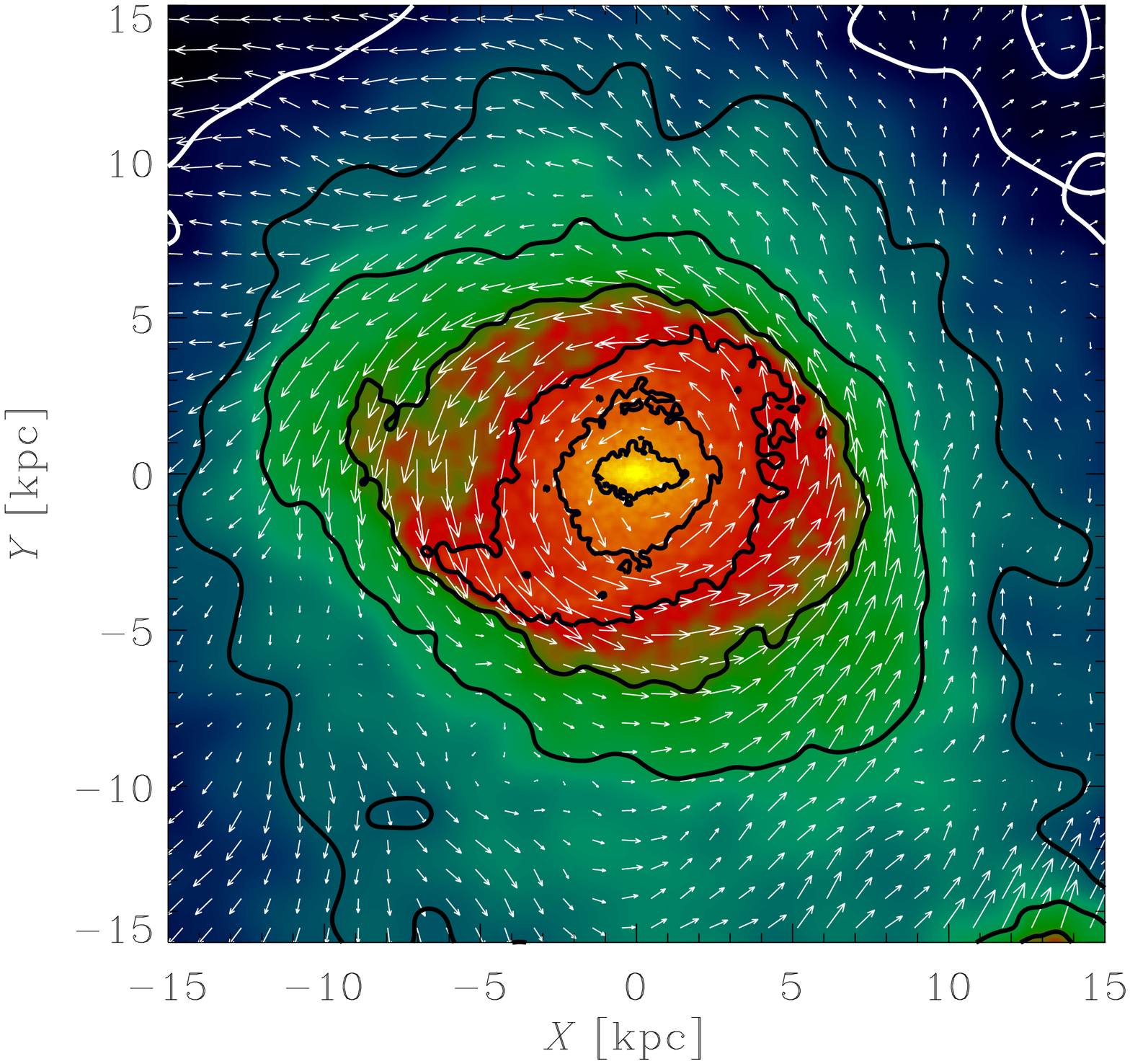}\includegraphics[height=3.2cm]{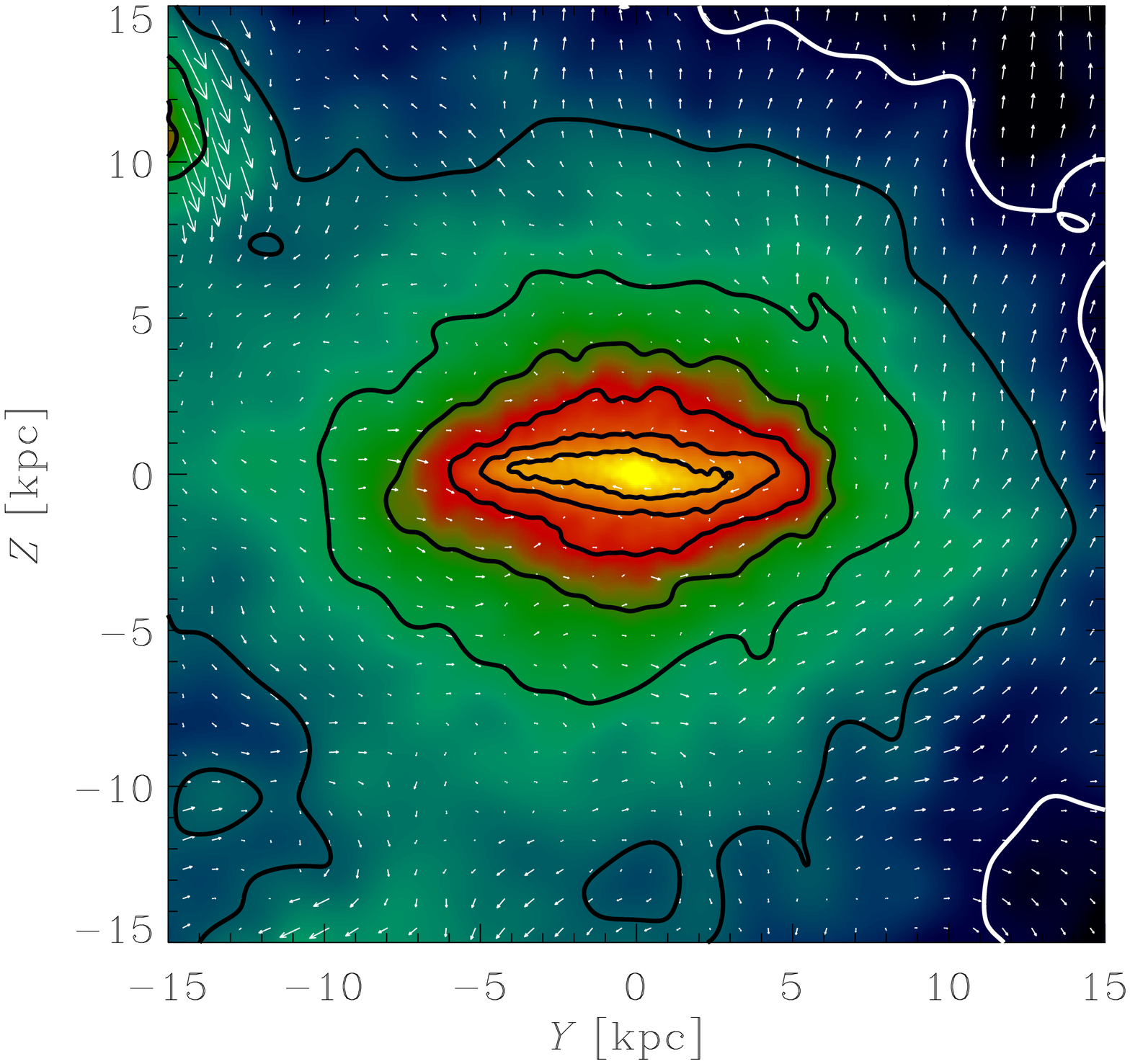}\hspace{1cm}\includegraphics[height=3.2cm]{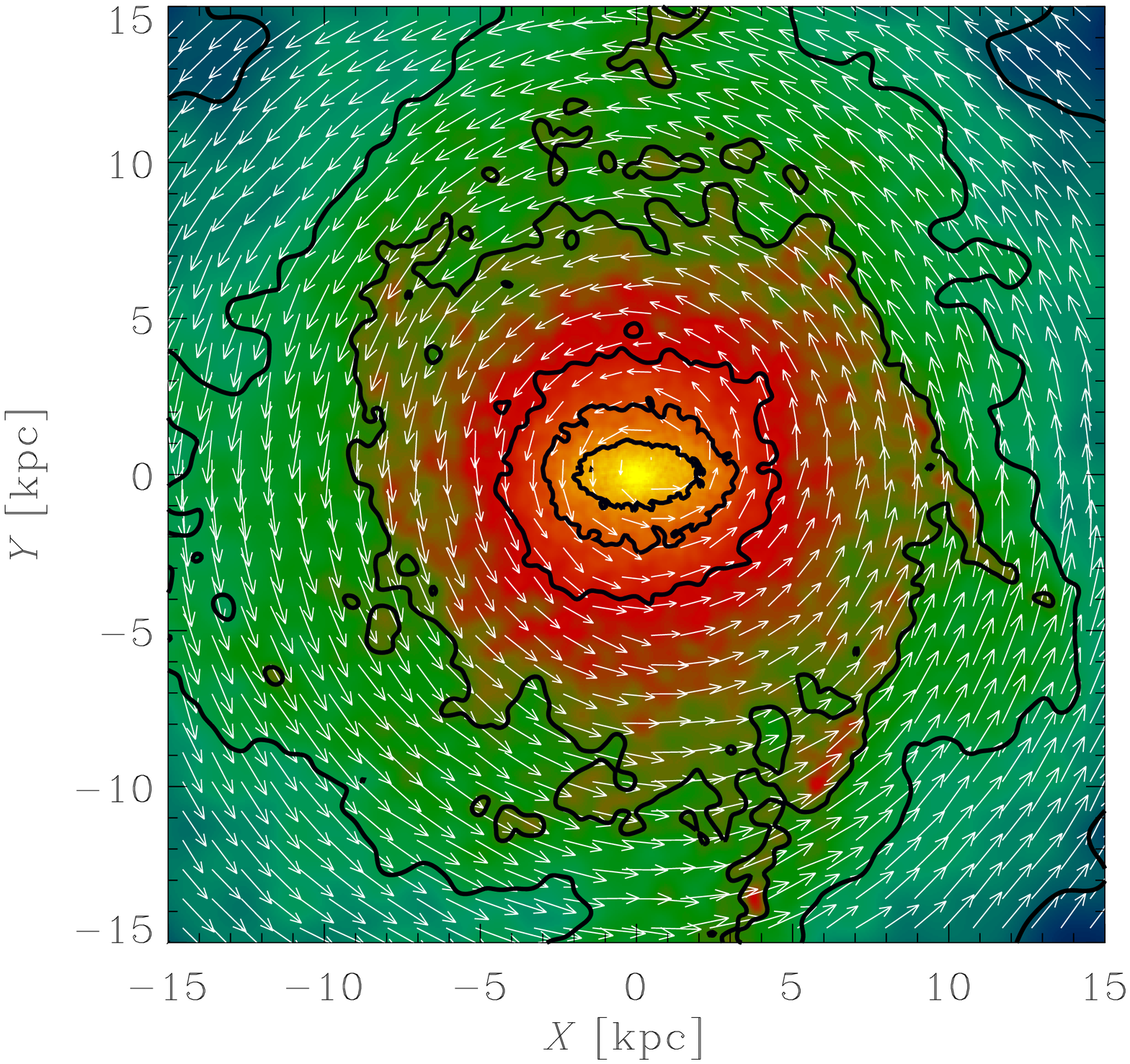}\includegraphics[height=3.2cm]{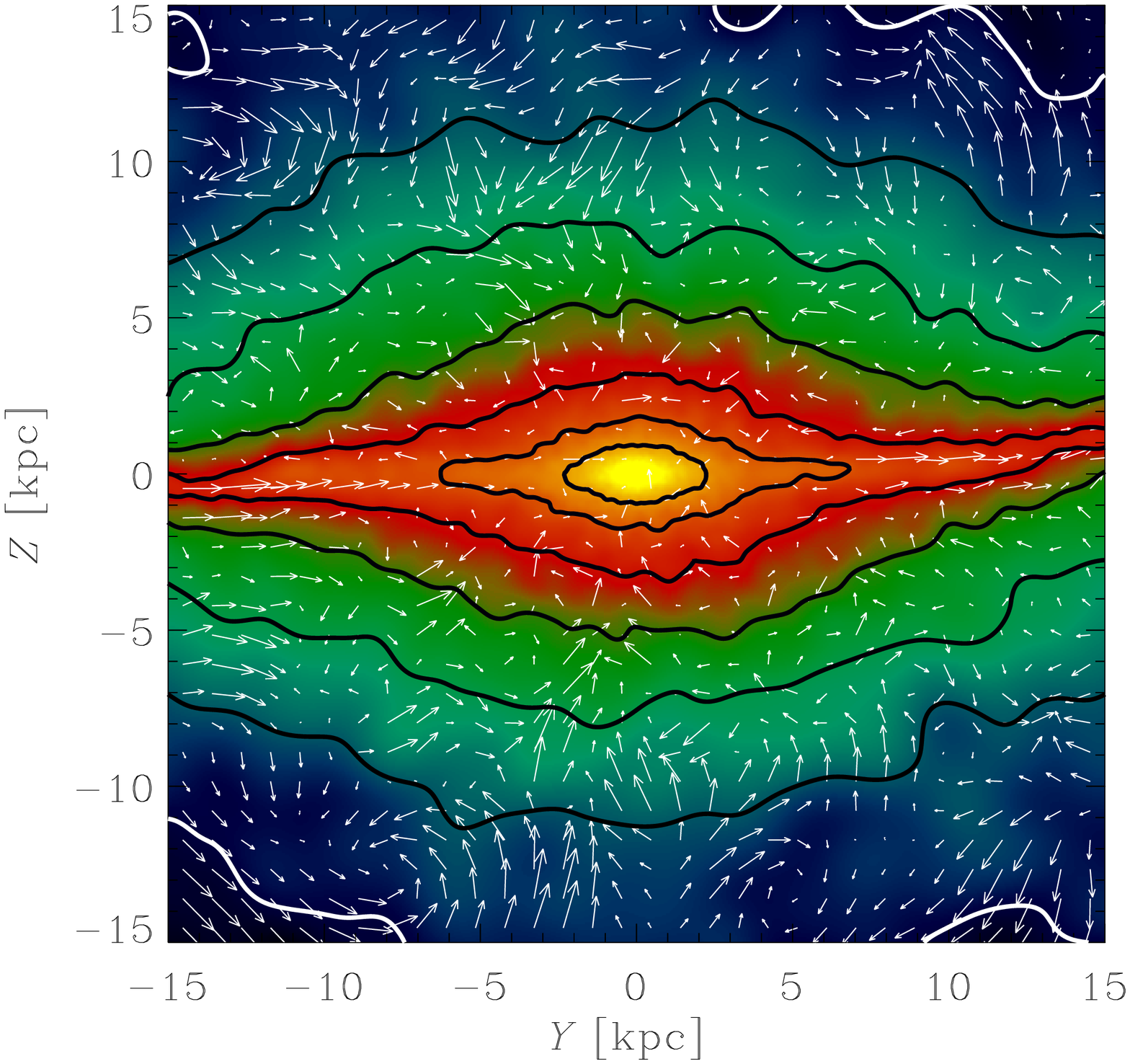}}

\caption{Projected stellar mass density of G1 (left-hand panels) and G2 (right-hand panels) 
         showing edge-on and face-on views for simulation runs CS (upper panels) and MA (lower panels). 
          Also shown are the velocity field (arrows) and isodensity contours 
         (solid lines) in the corresponding plane. 
         }
\label{fig:G1_G2_z0}
\end{center}
\end{figure*}

\begin{figure*}
\begin{center}
\includegraphics[width=15cm]{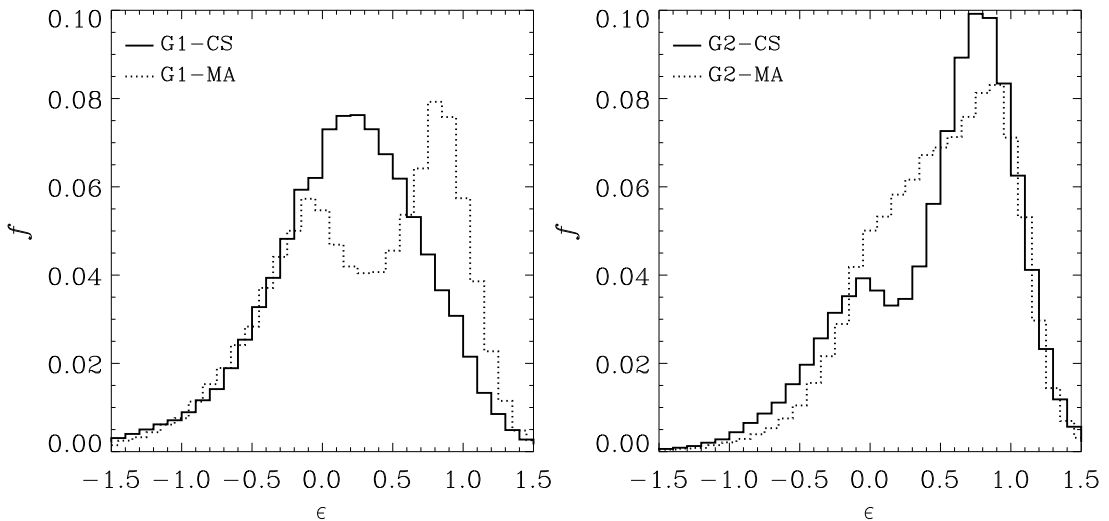}
\caption{Distribution of stellar circularities of G1 (left-hand panel) and G2 (right-hand panel) 
         for simulation runs CS (solid lines) and MA (dotted lines) at $z=0$.}
\label{fig:circ}
\end{center}
\end{figure*}

The stellar distributions of the central, main component of G1 and G2 
in our two runs and at $z=0$ are
shown in  Fig.~\ref{fig:G1_G2_z0}
for face-on 
and edge-on views, in an area of ($40$ kpc)$^2$.
To create these maps, we have rotated the galaxies
such that the stellar angular momentum within 
a  radius of
$30$ ckpc lies in the $z-$direction. 
We also plot isodensity contours at fixed density values
in order to highlight differences in the mass distributions 
of the two simulations. 

\begin{figure*}
\begin{center}
\includegraphics[width=17cm]{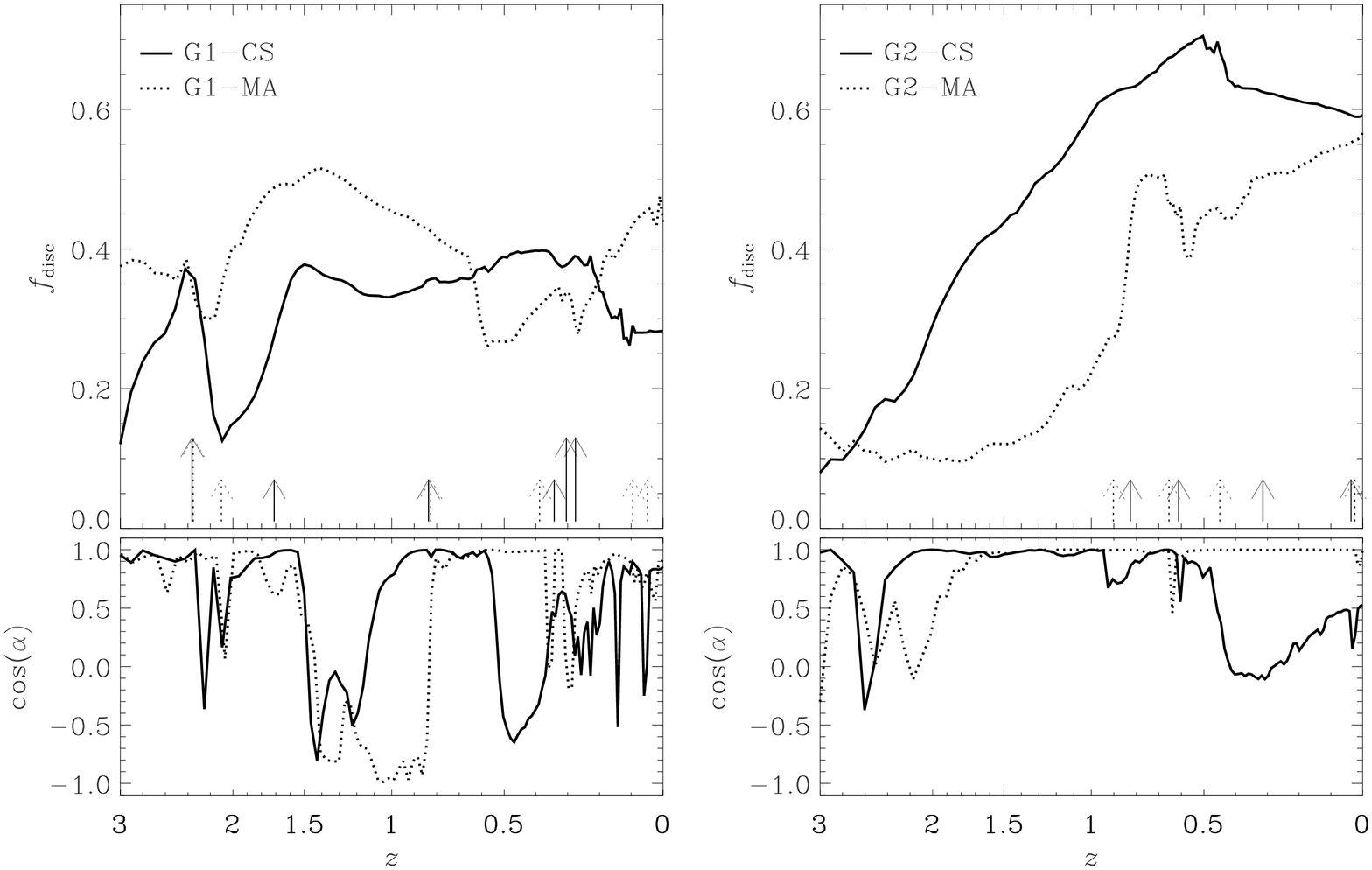}

\caption{ The evolution of the disc-to-total measure $\fdisc$ 
 of G1 and G2 in our two simulations. The long/short arrows denote
the entrance of major/intermediate mass mergers in the inner 30 ckpc in the
two runs. 
The lower panels show the  evolution of cos($\alpha$), the angle between
the corresponding angular momentum vectors of the gas and stars. }
\label{fig:dt}
\end{center}
\end{figure*}

Immediately apparent is that, in both runs,  G1 is a more spheroidal
galaxy than G2, which has a well-defined, extended,
disc-like component.
The presence/absence of a disc can be inferred from the projected
velocity fields overplotted to the stellar densities. These
show that G1 has a certain degree of rotational
support that is however not very significant.
Compared to
G1-CS,  the rotation in G1-MA is much more coherent,
as  evidenced not only by the tangential velocities
in the face-on view but also by the low velocities observed in the edge-on projection.
A much higher degree of rotation is found for G2 in both
simulations, where a well-defined, extended disc component is present. 
As shown later (Figs.~\ref{fig:circ} and ~\ref{fig:dt}) 
neither G2-CS nor G2-MA has a dominant, massive bulge in the centre, and 
neither shows evidence of  a bar. Note, however,
that due to its younger age (Fig.~\ref{fig:dt}, see below), 
the disc of G2-MA is thinner
compared to that of G2-CS.
Also note that, both for G1 and G2,  we find that the stellar haloes 
are more extended in run CS compared to run MA, which is
a direct consequence of the stronger feedback assumed in the latter which
prevents the formation of the old stars that make up 
this component.

In order to better quantify the dynamical state of G1 and G2
and compare results from our two simulations, we show in
 Fig.~\ref{fig:circ} the corresponding distributions  of stellar circularities.
The circularity of a star $i$ is defined as in \cite{S09}:
\begin{equation}\label{eq:epsilon}
\epsilon_i \equiv {j_z^i\over{j_{\rm circ}(r_i)}} 
\end{equation}
where $j_z^i$ is the angular momentum of the $i$ star particle 
perpendicular to the disc
plane (i.e. in the $z-$direction), 
and $j_{\rm circ}$ the angular momentum expected for a circular orbit at
the star's radius. Circularities of the order of 1 are expected 
for rotationally-supported particles while
spheroidal components form more or less symmetric distributions around
$\epsilon=0$ (non-rotating bulges), $0\lesssim\epsilon\lesssim 0.5$ (rotating
bulges) or  $-0.5\lesssim\epsilon\lesssim0$ (counter-rotating bulges).

As inferred from Fig.~\ref{fig:G1_G2_z0},  Fig.~\ref{fig:circ} confirms 
that G1 has an important spheroidal-like component. In G1-CS,
the spheroid has an overall rotation (the peak of
the distribution is at positive $\epsilon$), while in G1-MA the spheroid
does not rotate. Furthermore, G1-MA has a disc-like component in rotational
support.
 As we show later, the presence/absence of the disc in G1-CS
relates to the fact that this galaxy has experienced, before $z=0$, 
a number of mergers which destroyed a pre-existing disc, while these
events have not yet happened in G1-MA at $z=0$. 
In the case of
G2, the presence of a well-defined, rotationally-supported disc is clear in both
runs. In addition, G2-CS has  a non-rotating, distinguishable
although not-dominant bulge.

\begin{figure*}
\begin{center}
{\includegraphics[width=9cm]{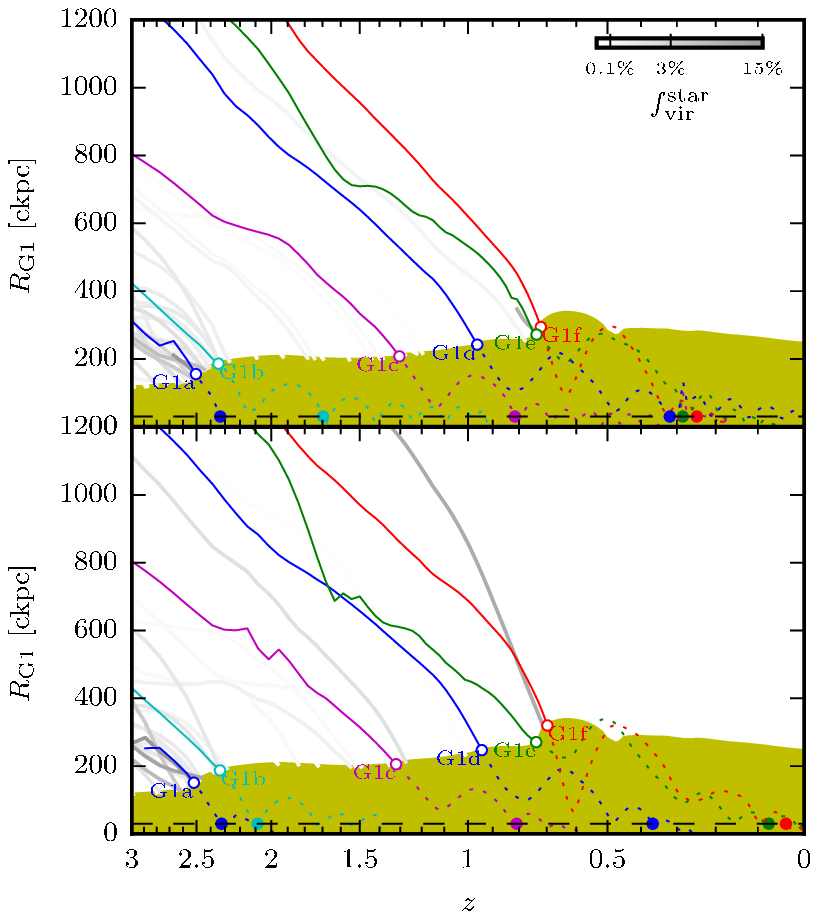}\includegraphics[width=9cm]{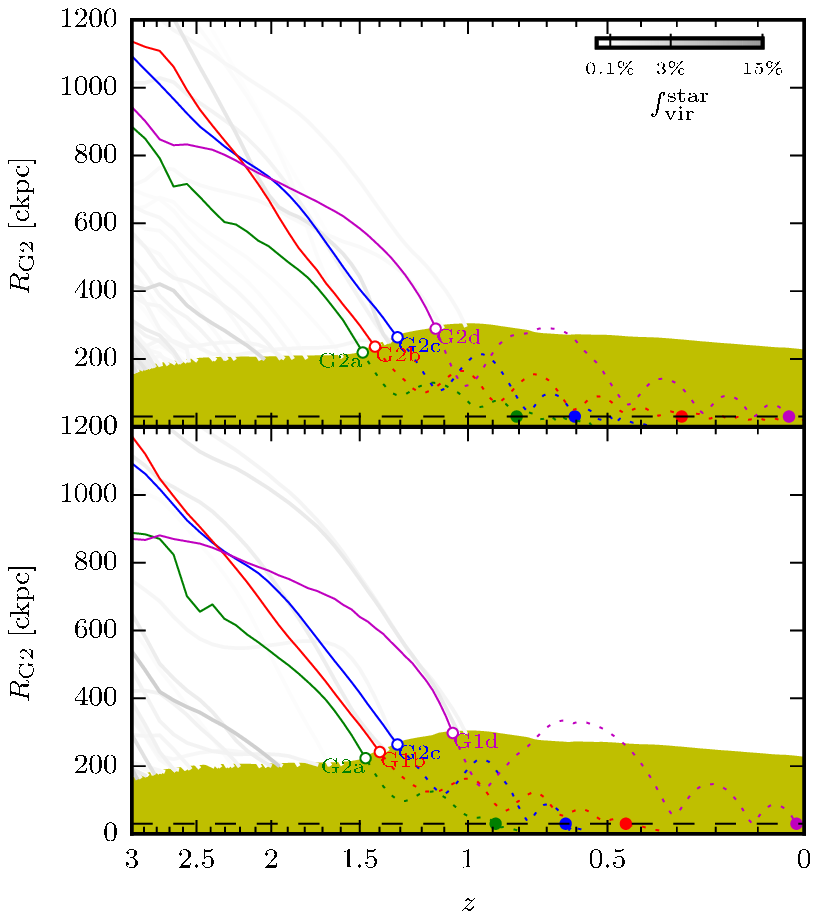}}
\caption{The merger history of G1 (left-hand panels) and G2 (right-hand
panels) for simulations CS (upper panels) and MA (lower panels), 
alongside the evolution of its statistics. 
The plots show the comoving distance of each satellite 
from the position of G1/G2 as a function of redshift, 
where \emph{coloured lines} indicate the major ($> 30 \%$) and 
 intermediate mass  (between $5$ and $30\%$) mergers (see
Table~\ref{tab:merger_prop}). The greyscales indicate minor 
mergers, the tone representing $f^{\rm star}_{\rm 200}$, the stellar mass of the 
progenitor as a fraction of the stellar mass of the galaxy 
at the time of first infall into the corresponding 
$R_{200}$ (see Section \ref{sec:mergers} of the main text for full details). 
The   \emph{shaded region} indicates $R_{200}$ of G1/G2, 
whilst the \emph{black horizontal dashed line} indicates a radius $30$ ckpc, a cut which encompasses the baryonic component  of the galaxy. 
The \emph{dotted lines} follow the positions of the major progenitors after first 
infall within the $R_{200}$  until {\sc subfind} can no longer identify them as distinct objects. }
\label{fig:mergertrees}
\end{center}
\end{figure*}

The SFRs of G1 and G2 in our two simulations,  shown in the previous section
(Fig.~\ref{fig:sfr}) and, in particular, the corresponding in-situ fractions,
already gave some indications of what the final morphology
of the galaxy would be. In G1, star formation is extremely bursty,
and the in-situ fractions are relatively low, which is indicative of a large contribution
of stars from satellite galaxies. 
In general such a contribution from satellites will reduce
the disc fraction, unless the angular momentum of the satellite
is very closely aligned with that of the progenitor galaxy.
In the case of G2, the smoother SFRs
together with very high in-situ fractions, particularly
at low redshifts, suggest the formation of a galaxy
with a much more significant disc which did not suffer
disruption from violent events.

In the rest of this section we show that, in fact, G1 has a much more
violent
merger history than G2 which has a much
smoother evolution. 
It is worth noting that the in-situ fractions of our simulated
galaxies are similar in runs CS and MA
which is
 an indication that our results on the final morphologies 
(particularly of G1) are robust and not triggered
by the assumed feedback model. Mergers seem to be as disruptive in
both runs, regardless differences in the exact merging times of the
satellites (see below)
and the different final stellar masses of the galaxies.

Fig.~\ref{fig:mergertrees} shows the merger histories of
G1 and G2 in our two simulations, 
which we constructed 
by identification of haloes using the substructure
finder code
{\sc subfind} 
\citep{Springel_2001} and then matched between snapshots with a metric
that weights intersecting particles by the ranking of their binding energy. 
We identified all mergers with a mass ratio larger than $0.1\%$ in terms of their total mass, 
and denoted any merger
greater than $30\%$ as a `major' merger and any greater than $5\%$ as an `intermediate'  merger. 
Table \ref{tab:merger_prop} describes statistics -- redshift and mass ratio --  of the 
intermediate and major mergers as the satellites fall within the $R_{200}$  of the progenitor galaxy.
We also show the  mass ratios in terms of the
stellar mass, as these might be more indicative of the importance of mergers
in the central regions.

Whilst the accretion of a progenitor into $R_{200}$ will clearly play a role 
in the evolution of the galaxy, the events at much smaller radii have a more 
direct impact, as an infalling object can orbit the galaxy for several Gyrs 
(or more) before merging into the central object. For this reason we followed
the merging subhaloes also after their entrance into $R_{200}$,
and calculated the times and mass fractions when they enter the inner 30 ckpc,
where we study the stellar 
and gaseous components. This radius is a good choice as it is both 
large enough to encompass the galaxy (see Fig.~\ref{fig:G1_G2_z0})
but also small enough that objects whose centres have passed this close will significantly interact. 
For both simulations, the last three columns of
Table \ref{tab:merger_prop} show the redshift at which satellites
cross the 30 ckpc  threshold, as well as the stellar mass ratio at that time.
Note that,
in order to match results of runs CS and MA, we followed
the mergers of satellite galaxies even though, in some cases, they
have not yet merged/disrupted with the main progenitor
and are still identifiable, independent systems at the present time.
These are indicated  in the final column of the
Table~\ref{tab:merger_prop} with annotation `sat'.
Also note that the times of the mergers, particularly when the satellites
reach the inner 30 kpc, are different in the two runs, as
well as the merger ratios.

From Fig.~\ref{fig:mergertrees} and Table~\ref{tab:merger_prop} we can see 
that, as inferred from the growth of the haloes and the
in-situ fractions, G1 has a more active history than G2,
with 6 major/intermediate
mass mergers, compared to only 4 in the case of G2.
Perhaps more significant is that
at late times ($z<1$) G1 accretes three very large objects indeed,
while for G2 there are no mergers in this period (considering their
entrance to $R_{200}$),
and its only mergers for $z \in [0,3]$ are at much 
more modest stellar fractions.
Although similar in general terms, the details of the merging
process of G1 and G2 in our two runs\footnote{Note that this can explain
some of the differences in the properties of G1 and G2 in runs CS and MA, e.g.
morphologies,  that
we discussed in this and the previous sections.} show some important differences.
In particular, in G1-CS only one
(and small) satellite survives at $z=0$
as an independent object, while the same object has been
already disrupted in run MA. More important is the fact that
the two satellites which merge later (denoted G1e/f) 
have been disrupted by $z=0$ in run CS but not yet in run MA. 
For G2, we find less variation between the simulations
and, in any case, the merger events of G2 are at
much lower fractions and occur earlier compared to
those of G1. For this reason, it is expected that the 
merger history will have a much stronger impact in the
formation of G1 compared to G2.

\begin{table*}
\caption{Properties of the major and intermediate mass mergers of G1 and G2
in our two simulations. The G1a-f denotes these 5 for G1, and G2a-d the 2 for G2, 
ordered by their appearance within the 30 ckpc. $z_{30\, \rm ckpc}$ and 
$z_{\rm 200}$ denote the highest redshift at which the most bound 
{\sc subfind} particle enters within 30 ckpc  and $r_{\rm 200}$ respectively. 
$f^{\rm star}_{30 \, \rm ckpc}$ and $f^{\rm star}_{\rm R200}$ denote the {\sc subfind} 
stellar mass of the satellite as a fraction of that of the main galaxy at the corresponding redshift. 
$f^{\rm tot}_{\rm R200}$ is the total mass fraction of the satellite when it reaches $R_{200}$. 
For the most massive mergers (G1a, G1d, G1e, G2b) the {\sc subfind} mass estimate exhibits
significant disruption as the object enters $R_{200}$, and so for these cases 
we calculate the ratios 200 Myr earlier to avoid spurious values. In the final column,
we indicate with `sat' those satellite galaxies that have not merged/disrupted at $z=0$.
}
\begin{center}
\begin{tabular}{ l | rrrrrc | lrrrrrc}
 
& \multicolumn{6}{c}{Run CS} &  \multicolumn{6}{c}{Run MA} \\
 &  $z_{\rm R200} $  & $f^{\rm tot}_{\rm R200}$  & $f^{\rm star}_{\rm R200}$ & $z_{30 \rm \, ckpc} $ & $f^{\rm star}_{30 \, \rm ckpc}$  & $z=0$ &   $z_{\rm R200} $  & $f^{\rm tot}_{\rm R200}$  & $f^{\rm star}_{\rm R200}$&$z_{30 \rm \, ckpc} $ & $f^{\rm star}_{30 \, \rm ckpc}$ &$z=0$\\\\
\hline\\
G1a  & $2.52$ & $67\%$ &$51\%$ & $2.33$ & $21\%$& & $2.52$ & $121\%$&$48\%$ &$2.32$ & $111\%$&\\
G1b & $2.35$ & $5\%$ &$8\%$    & $1.70$ & $2\%$ &  & $2.35$ & $6\%$  &$8\%$ &$2.09$ & $5\%$ &\\
G1c & $1.29$ & $8\%$&$6\%$      & $0.82$ & $5\%$&  & $1.33$ & $12\%$ &$10\%$&$0.81$ & $14\%$ &\\
G1d  & $0.95$ & $13\%$&$14\%$   & $0.32$ & $5\%$& sat & $0.95$ & $13\%$ &$33\%$&$0.37$ & $28\%$ &\\
G1e & $0.74$ &$63\%$  &$44\%$  & $0.28$ & $27\%$&   & $0.74$ & $15\%$ &$80\%$&$0.08$ & $21\%$& sat\\
G1f   & $0.72$ & $64\%$ &$78\%$& $0.25$ & $36\%$&  & $0.70$ & $19\%$ &$108\%$&$0.04$ & $43\%$ & sat\\
\\\hline\\
G2a     & $1.48$ & $11\%$ & $11\%$& $0.81$ & $5\%$&    & $1.48$ & $13\%$ &$29\%$&$0.89$ & $16\%$ &\\
G2b     & $1.44$ & $9\%$ & $10\%$ & $0.29$ & $2\%$& sat & $1.40$ & $13\%$  &$33\%$&$0.44$ & $9\%$&\\
G2c   & $1.33$ & $15\%$ & $28\%$& $0.60$ & $9\%$  &     & $1.33$ & $18\%$ &$61\%$&$0.64$ & $33\%$&\\
G2d    & $1.14$ & $7\%$ & $14\%$ & $0.03$ & $5\%$ &sat & $1.05$ & $8\%$ &$37\%$&$0.02$ & $9\%$ &sat\\
\end{tabular}
\label{tab:merger_prop}
\end{center}
\end{table*}

The growth/destruction of galaxy discs is intimately related to
the merger histories of galaxies. In fact, we find
a clear relation between periods of disc growth/destruction
and the merger history of our simulated galaxies. 
Fig.~\ref{fig:dt}
shows the evolution of $\fdisc$, a measure of the disc-to-total (D/T) ratio,
for G1 and G2 in runs CS and MA. Note that, 
from Fig.~\ref{fig:circ}, it is clear that the distribution
of circularities is a very clear indicator of the presence or absence of a
disc-like stellar component and thus a single quantitative measure of this is the 
fraction of stellar mass with circularity larger\footnote{In the calculation of $f_{\rm disc}$, we consider only 
particles of the main subhalo, excluding satellites. } than 
0.5,
\begin{equation}
\fdisc\equiv f(\epsilon>0.5) \, .
\end{equation}
Note, however,  that $\fdisc$ can be non zero even if there is no disc,
if the distribution of circularities is sufficiently broad.
On the other hand, previous work  (\citealt{S10}; 
see also \citealt{Abadi03}, \citealt{Governato07}) showed  that 
kinematic D/T decompositions, as the one we present here, 
give ratios that are systematically and significantly lower than those
obtained with photometric decompositions
(i.e. comparable with observed D/T ratios). For example, in \cite{S10} 
we showed  that D/T can increase from $0.2$ (in a kinematic decomposition)
up to $0.5-0.7$ (in a photometric decomposition).
In any case, as the primary focus of our analysis is to find an 
appropriate measure of the evolution of a galaxy's morphology
(and not to make a detailed comparison of simulated and observed
D/T ratios), we can safely use  $\fdisc$ as it 
sufficiently captures the prominence of the disc as a function of time in our
galaxies.

The more violent merger history of G1 is also clear from the evolution
of $\fdisc$, as  G1 exhibits significant morphological evolution 
between $z=3$ and $z=0$.  Most of the rotational support
of G1 is acquired between $z\sim 2$ and $z\sim 1.5$ in both
runs. In G1CS, the disc is stable until $z\sim 0.3$, when 
it is disrupted coinciding with the merger
events occurring at $z\sim$$~0.3-0.1$.
For G1-MA, we find an important
decrease in $\fdisc$ between $z=1.5$ and $z\sim0.5$,
and a second phase of disc growth after
$z\sim 0.4$. Despite the differences in the evolution
of G1-CS and G1-MA, mainly driven by the different
merger times of the satellite galaxies (when they enter the inner
regions), in
both simulations we find a galaxy where the rotational
support, when present, cannot be maintained for long periods
of time.

The more peaceful merger history of the G2 halo is consistent 
with the rearing of a more disc-dominated galaxy,
and a more continuous and smooth disc growth. 
In both simulations, G2 evolves into a galaxy with a significant
disc, little affected by disruptive merger events.
An important difference is however the typical age and formation
time of the discs
in the two simulations: the disc of G2-CS is present
from very early times, and therefore is older on average, while
in G2-MA the disc does not form until after $z=1$ and never
reachs  $\fdisc$ values as large as those found for G2-CS.
This is, as explained before, the result of the stronger
feedback assumed in the MA model, which produces a lower
star formation activity at early times due to the 
prompt effects of radiation pressure  and kinetic feedback
from young stars.

Finally,  in the lower panels of Fig.~\ref{fig:dt}
we show the corresponding evolution of the cosine of the angle $\alpha$ between
the specific angular momenta of the gas (${\bf j}_{\rm gas}$) and 
of the stars (${\bf j}_{\rm stars}$), namely
\begin{equation}
{\rm cos}(\alpha) \equiv {\bf{\hat{j}}}_{\rm stars}\cdot{\bf{\hat{j}}_{\rm gas}} \, ,
\end{equation}
where the hat 
denotes a normalised vector. 
In previous work, we  found that
if the angular momentum vectors of the gas and the stars are misaligned,
the stellar discs become unstable and the D/T ratios inevitably decrease \citep{S09}.
Such a misalignment can occur as a consequence of merger events,
but also if accreting gas brings angular momentum in a  different direction
than that of the pre-existing stellar disc.

Consistent with our previous analysis, we find
that the alignment between the gas and stars is not well
preserved during the evolution of G1, mainly as a consequence
of the major merger events of this galaxy, which
 determine  the end points of the growth of the discs. The results
are very similar for our two runs, which is again an indication
that results are robust and not affected by the particular
feedback assumed  but rather by the merger history.
In the case of G2, the gas and the stars are much
better aligned at all times, in both runs. The most important difference
in this case is at late times, when G2-CS shows a strong
misalignment, consequence of the mergers that are
taking place at $z\lesssim 0.5$. As seen before, in G2-MA 
one of such
satellites has already been disrupted while the other has not yet 
reached the central object.

In summary, by comparing these two histories we see a picture where the 
discs
are built-up during periods of absence of major mergers
and alignment between the gaseous and stellar discs. 
In both simulated galaxies the growth of the discs suffers 
termination events when such alignment is disrupted, usually  
associated with merging satellites (but can also be produced by misaligned gas
accretion), either when they pass within 30 ckpc 
or merely have 
a close approach.

\section{Discussion and conclusions}\label{sec:conc}

We used a simulation of the Local Universe
where a Local Group-like pair of galaxies with large-scale
environment constrained to reproduce the environment of the actual
LG forms. We study how the two most massive galaxies within the 
simulated LG, 
candidates for the Milky Way and Andromeda, formed and evolved.
The constrained nature of the simulation is relevant \citep{Creasey15},
as it allows the study of the evolution of galaxies in an environment
similar to our LG, and thus the investigation of  how it affects
the properties of the galaxies along cosmic time.

We run two simulations using the same initial conditions, but
 assuming different feedback prescriptions, in
order to assess whether our results might in any sense
be affected by the particular choice of feedback model. 
It is well known that the detailed properties of simulated
galaxies depends on the implementation of the physical processes
included, in particular on the way stellar feedback
is treated \citep{S12}. Furthermore, it is still not
possible to select the several input parameters of the
simulation from physical principles, and various numerical/technical
choices can also affect the final outcome of simulations.
Despite these problems, which will certainly
witness significant progress in the next few years,
the predictions of simulations
in relation to the {\it evolution} of properties
such as  morphologies should be still reliable.

In our simulations, 
the two main galaxies of the simulated LG, G1 (candidate for M31)
and G2 (candidate for MW), 
exhibit rather different formation and merger histories
although they are, at $z=0$, both members of the same LG. 
G1 is much more active in terms of the frequency and mass
ratios of
its mergers compared to G2. More importantly,
G1 experiences three major/intermediate mass mergers at $z<1$, contrary to
G2 that has (since $z=3$) only four intermediate mass mergers,
and these occur before $z=1$.
The merger rates of the two galaxies are then a first indication
that the two galaxies will follow different evolutionary paths,
in particular in relation to
the formation/destruction of stellar discs.

In fact, we find that G1, having 6 intermediate/major mergers
between $z=3$ and $z=0$, cannot grow a stellar disc that survives
until the present time. The various mergers
disturb G1 significantly, changing the amount and  orientation
of the angular momentum of the stars and gas, that in both codes leads to a 
posterior misalignment of the two vectors.
When such a misalignment is present, any pre-existing disc
gets fully or partially destroyed. 
In contrast, the much more smooth merger history
of G2, where
the angular momentum of the 
stars and gas are very well aligned during most of the time,
is consistent with the formation of an extended, rotationally supported
disc-like galaxy. 
It is worth noting that mergers, even fairly sizeable ones, do not
necessarily destroy discs, and the actual damage depends on
the orbital parameters (see also \citealt{Wang12}).

The different merger/formation histories of G1 and G2 is also reflected
in their star formation rates and in-situ fractions. G1 has a much more violent
SFR, and low in-situ fractions, indicating a large
contribution of stellar material from infalling satellites. With such 
an active merger history, it seems very hard to grow
a stellar disc with the satellites feeding the disc,
instead
of destroying it. The more peaceful history of G2
and its much  higher in-situ fractions at all times
suggest that this galaxy will have a much
higher chance of disc survival.

We investigated and compared the evolution of the MW and M31 candidates in
two runs, 
a first one where only supernova feedback is considered (CS) and
a second one that additionally includes the effects
of radiation pressure from massive stars and a kinetic component
for the supernova energy. Such comparison allowed us to assess
the robustness of our results and their dependence on the
assumed feedback model.
While it is true that, due to the larger feedback effects, the MA model
produces galaxies with reduced stellar masses compared to CS, both for
the progenitor and for the satellites, 
mergers are as disruptive in the  MA model as
in the CS simulation, where galaxies have in general larger stellar masses. 
In both simulations we find that
the stellar disc of G1 suffers several episodes of
partial destruction that follow merger events, while 
in G2 a stellar disc grows steadily due to its more quiescent evolution.
We found (expected) 
differences in the exact merger times
of satellites between the two runs,
not only when they enter $R_{200}$  but more
importantly when they reach the inner galaxy.
These can cause some of the satellites to be, at
any given time, already disrupted  in one run but
not in the other. 
More important are the overall differences in
the star formation rates in terms of the assembly
of the stellar components: the galaxies in CS
are systematically older than in MA, which is
a direct result of the stronger feedback assumed
in the latter.

In summary, by comparing the histories of formation of G1 and G2, and its relation
to the formation/destruction of stellar discs along cosmic time, 
we conclude that the rearing of a disc can only be
produced during quiet periods where no significant mergers and/or
close encounters are produced,
and when the angular momentum vectors of the stars and the gas are
well aligned. These results are consistent with those found earlier
in S09, where we simulated the formation of
eight galaxies in a more isolated environment,
giving support to the idea that it is the particular
formation history of haloes at relatively small scales (smaller
at least than the LG) that determines
their morphological changes along cosmic time, rather
than their membership of a loose group such as the LG.

\begin{acknowledgements}

We thank the reviewer for  his/her constructive suggestions. 
We thank M. Aumer for fruitful discussions on the implementation
of radiation pressure and kinetic feedback on the simulation code. 
The simulation was performed on the Juropa supercomputer of 
the J\"ulich Supercomputing Centre (JSC), and initial tests
were run at the Barcelona Supercomputer Centre (BSC). 
C.S. and P.C. acknowledge support from the Leibniz Gemeinschaft through
grant SAW-2012-AIP-5 129. 
C.S.  acknowledges support from the HPC-EUROPA2 project, with the support of the European Community - Research Infrastructure Action of the FP7.
S.E.N. acknowledges support from the Deutsche Forschungsgemeinschaft under the 
grants MU 1020 16-1 and NU 332/2-1, and 
G.Y.   thanks  MINECO  (Spain)  for supporting his research through different  projects:  AYA2012-31101, FPA2012-34694 and Consolider Ingenio SyeC CSD2007-0050.
\end{acknowledgements}

\bibliographystyle{mn2e}

\bibliography{biblio.bib}

\begin{thebibliography}{55}
\expandafter\ifx\csname natexlab\endcsname\relax\def\natexlab#1{#1}\fi

\bibitem[{{Abadi} {et~al}\mbox{.}(2003){Abadi}, {Navarro}, {Steinmetz}, \&
  {Eke}}]{Abadi03}
{Abadi} M.~G., {Navarro} J.~F., {Steinmetz} M., {Eke} V.~R., 2003, ApJ, 591,
  499

\bibitem[{{Agertz} \& {Kravtsov}(2014)}]{Agertz15}
{Agertz} O., {Kravtsov} A.~V., 2014, ArXiv e-prints (1404.2613)

\bibitem[{{Agertz} {et~al}\mbox{.}(2013){Agertz}, {Kravtsov}, {Leitner}, \&
  {Gnedin}}]{Agertz13}
{Agertz} O., {Kravtsov} A.~V., {Leitner} S.~N., {Gnedin} N.~Y., 2013, \apj,
  770, 25

\bibitem[{{Aumer} {et~al}\mbox{.}(2013){Aumer}, {White}, {Naab}, \&
  {Scannapieco}}]{Aumer13}
{Aumer} M., {White} S.~D.~M., {Naab} T., {Scannapieco} C., 2013, \mnras, 434,
  3142

\bibitem[{{Bah{\'e}} {et~al}\mbox{.}(2013){Bah{\'e}}, {McCarthy}, {Balogh}, \&
  {Font}}]{Bahe_2013}
{Bah{\'e}} Y.~M., {McCarthy} I.~G., {Balogh} M.~L., {Font} A.~S., 2013, \mnras,
  430, 3017

\bibitem[{{Bird} {et~al}\mbox{.}(2013){Bird}, {Kazantzidis}, {Weinberg},
  {Guedes}, {Callegari}, {Mayer}, \& {Madau}}]{Bird13}
{Bird} J.~C., {Kazantzidis} S., {Weinberg} D.~H., {Guedes} J., {Callegari} S.,
  {Mayer} L., {Madau} P., 2013, \apj, 773, 43

\bibitem[{{Blanton} {et~al}\mbox{.}(2003){Blanton}, {Eisenstein}, {Hogg},
  {Schlegel}, {Brinkmann}, {Quintero}, {Berlind}, \& {Wherry}}]{Blanton03}
{Blanton} M.~R., {Eisenstein} D.~J., {Hogg} D.~W., {Schlegel} D.~J.~S.,
  {Brinkmann} J., {Quintero} A.~D., {Berlind} A., {Wherry} N., 2003, in
  Bulletin of the American Astronomical Society, Vol.~36, American Astronomical
  Society Meeting Abstracts, p. 145.01

\bibitem[{{Bovy} {et~al}\mbox{.}(2012){Bovy}, {Allende Prieto}, {Beers},
  {Bizyaev}, {da Costa}, {Cunha}, {Ebelke}, {Eisenstein}, {Frinchaboy},
  {Garc{\'{\i}}a P{\'e}rez}, {Girardi}, {Hearty}, {Hogg}, {Holtzman}, {Maia},
  {Majewski}, {Malanushenko}, {Malanushenko}, {M{\'e}sz{\'a}ros}, {Nidever},
  {O'Connell}, {O'Donnell}, {Oravetz}, {Pan}, {Rocha-Pinto}, {Schiavon},
  {Schneider}, {Schultheis}, {Skrutskie}, {Smith}, {Weinberg}, {Wilson}, \&
  {Zasowski}}]{Bovy12}
{Bovy} J. {et~al.}, 2012, \apj, 759, 131

\bibitem[{{Christensen} {et~al}\mbox{.}(2012){Christensen}, {Quinn},
  {Governato}, {Stilp}, {Shen}, \& {Wadsley}}]{Christensen12}
{Christensen} C., {Quinn} T., {Governato} F., {Stilp} A., {Shen} S., {Wadsley}
  J., 2012, \mnras, 425, 3058

\bibitem[{{Corbelli} {et~al}\mbox{.}(2010){Corbelli}, {Lorenzoni}, {Walterbos},
  {Braun}, \& {Thilker}}]{Corbelli10}
{Corbelli} E., {Lorenzoni} S., {Walterbos} R., {Braun} R., {Thilker} D., 2010,
  \aap, 511, A89

\bibitem[{{Creasey} {et~al}\mbox{.}(2015){Creasey}, {Scannapieco}, {Nuza},
  {Yepes}, {Gottl{\"o}ber}, \& {Steinmetz}}]{Creasey15}
{Creasey} P., {Scannapieco} C., {Nuza} S.~E., {Yepes} G., {Gottl{\"o}ber} S.,
  {Steinmetz} M., 2015, \apjl, 800, L4

\bibitem[{{Creasey}, {Theuns} \& {Bower}(2013){Creasey}, {Theuns}, \&
  {Bower}}]{Creasey13}
{Creasey} P., {Theuns} T., {Bower} R.~G., 2013, \mnras, 429, 1922

\bibitem[{{Dressler}(1980)}]{Dressler80}
{Dressler} A., 1980, \apj, 236, 351

\bibitem[{{Fall} \& {Efstathiou}(1980)}]{Fall_Efstathiou_1980}
{Fall} S.~M., {Efstathiou} G., 1980, \mnras, 193, 189

\bibitem[{{Few} {et~al}\mbox{.}(2012){Few}, {Gibson}, {Courty},
  {Michel-Dansac}, {Brook}, \& {Stinson}}]{Few12}
{Few} C.~G., {Gibson} B.~K., {Courty} S., {Michel-Dansac} L., {Brook} C.~B.,
  {Stinson} G.~S., 2012, \aap, 547, A63

\bibitem[{{Girardi} {et~al}\mbox{.}(2003){Girardi}, {Rigoni}, {Mardirossian},
  \& {Mezzetti}}]{Girardi_2003}
{Girardi} M., {Rigoni} E., {Mardirossian} F., {Mezzetti} M., 2003, \aap, 406,
  403

\bibitem[{{Gottl\"ober}, {Hoffman} \& {Yepes}(2010){Gottl\"ober}, {Hoffman}, \&
  {Yepes}}]{Gottloeber10}
{Gottl\"ober} S., {Hoffman} Y., {Yepes} G., 2010, ArXiv e-prints

\bibitem[{{Governato} {et~al}\mbox{.}(2007){Governato}, {Willman}, {Mayer},
  {Brooks}, {Stinson}, {Valenzuela}, {Wadsley}, \& {Quinn}}]{Governato07}
{Governato} F., {Willman} B., {Mayer} L., {Brooks} A., {Stinson} G.,
  {Valenzuela} O., {Wadsley} J., {Quinn} T., 2007, MNRAS, 374, 1479

\bibitem[{{Guzzo} {et~al}\mbox{.}(1997){Guzzo}, {Strauss}, {Fisher},
  {Giovanelli}, \& {Haynes}}]{Guzzo97}
{Guzzo} L., {Strauss} M.~A., {Fisher} K.~B., {Giovanelli} R., {Haynes} M.~P.,
  1997, \apj, 489, 37

\bibitem[{{Haardt} \& {Madau}(1996)}]{HM96}
{Haardt} F., {Madau} P., 1996, ApJ, 461, 20

\bibitem[{{Hermit} {et~al}\mbox{.}(1996){Hermit}, {Santiago}, {Lahav},
  {Strauss}, {Davis}, {Dressler}, \& {Huchra}}]{Hermit96}
{Hermit} S., {Santiago} B.~X., {Lahav} O., {Strauss} M.~A., {Davis} M.,
  {Dressler} A., {Huchra} J.~P., 1996, \mnras, 283, 709

\bibitem[{{Hopkins} {et~al}\mbox{.}(2014){Hopkins}, {Kere{\v s}}, {O{\~n}orbe},
  {Faucher-Gigu{\`e}re}, {Quataert}, {Murray}, \& {Bullock}}]{Hopkins14}
{Hopkins} P.~F., {Kere{\v s}} D., {O{\~n}orbe} J., {Faucher-Gigu{\`e}re} C.-A.,
  {Quataert} E., {Murray} N., {Bullock} J.~S., 2014, \mnras, 445, 581

\bibitem[{{Kafle} {et~al}\mbox{.}(2012){Kafle}, {Sharma}, {Lewis}, \&
  {Bland-Hawthorn}}]{Kafle12}
{Kafle} P.~R., {Sharma} S., {Lewis} G.~F., {Bland-Hawthorn} J., 2012, \apj,
  761, 98

\bibitem[{{Kim} {et~al}\mbox{.}(2014){Kim}, {Abel}, {Agertz}, {Bryan},
  {Ceverino}, {Christensen}, {Conroy}, {Dekel}, {Gnedin}, {Goldbaum}, {Guedes},
  {Hahn}, {Hobbs}, {Hopkins}, {Hummels}, {Iannuzzi}, {Keres}, {Klypin},
  {Kravtsov}, {Krumholz}, {Kuhlen}, {Leitner}, {Madau}, {Mayer}, {Moody},
  {Nagamine}, {Norman}, {Onorbe}, {O'Shea}, {Pillepich}, {Primack}, {Quinn},
  {Read}, {Robertson}, {Rocha}, {Rudd}, {Shen}, {Smith}, {Szalay}, {Teyssier},
  {Thompson}, {Todoroki}, {Turk}, {Wadsley}, {Wise}, {Zolotov}, \& {AGORA
  Collaboration29}}]{AGORA}
{Kim} J.-h. {et~al.}, 2014, \apjs, 210, 14

\bibitem[{{Murante} {et~al}\mbox{.}(2015){Murante}, {Monaco}, {Borgani},
  {Tornatore}, {Dolag}, \& {Goz}}]{Murante15}
{Murante} G., {Monaco} P., {Borgani} S., {Tornatore} L., {Dolag} K., {Goz} D.,
  2015, \mnras, 447, 178

\bibitem[{{Navarro} \& {Benz}(1991)}]{Navarro91}
{Navarro} J.~F., {Benz} W., 1991, \apj, 380, 320

\bibitem[{{Nuza}, {Dolag} \& {Saro}(2010){Nuza}, {Dolag}, \& {Saro}}]{Nuza10}
{Nuza} S.~E., {Dolag} K., {Saro} A., 2010, \mnras, 407, 1376

\bibitem[{{Nuza} {et~al}\mbox{.}(2014{\natexlab{a}}){Nuza}, {Kitaura},
  {He{\ss}}, {Libeskind}, \& {M{\"u}ller}}]{Nuza14b}
{Nuza} S.~E., {Kitaura} F.-S., {He{\ss}} S., {Libeskind} N.~I., {M{\"u}ller}
  V., 2014{\natexlab{a}}, \mnras, 445, 988

\bibitem[{{Nuza} {et~al}\mbox{.}(2014{\natexlab{b}}){Nuza}, {Parisi},
  {Scannapieco}, {Richter}, {Gottl{\"o}ber}, \& {Steinmetz}}]{Nuza14}
{Nuza} S.~E., {Parisi} F., {Scannapieco} C., {Richter} P., {Gottl{\"o}ber} S.,
  {Steinmetz} M., 2014{\natexlab{b}}, \mnras, 441, 2593

\bibitem[{{Okamoto} {et~al}\mbox{.}(2005){Okamoto}, {Eke}, {Frenk}, \&
  {Jenkins}}]{Okamoto05}
{Okamoto} T., {Eke} V.~R., {Frenk} C.~S., {Jenkins} A., 2005, \mnras, 363, 1299

\bibitem[{{Piffl} {et~al}\mbox{.}(2014){Piffl}, {Scannapieco}, {Binney},
  {Steinmetz}, {Scholz}, {Williams}, {de Jong}, {Kordopatis}, {Matijevi{\v c}},
  {Bienaym{\'e}}, {Bland-Hawthorn}, {Boeche}, {Freeman}, {Gibson}, {Gilmore},
  {Grebel}, {Helmi}, {Munari}, {Navarro}, {Parker}, {Reid}, {Seabroke},
  {Watson}, {Wyse}, \& {Zwitter}}]{Piffl14}
{Piffl} T. {et~al.}, 2014, \aap, 562, A91

\bibitem[{{Piontek} \& {Steinmetz}(2011)}]{Piontek11}
{Piontek} F., {Steinmetz} M., 2011, \mnras, 410, 2625

\bibitem[{{Quinn}, {Hernquist} \& {Fullagar}(1993){Quinn}, {Hernquist}, \&
  {Fullagar}}]{Quinn93}
{Quinn} P.~J., {Hernquist} L., {Fullagar} D.~P., 1993, \apj, 403, 74

\bibitem[{{Sales} {et~al}\mbox{.}(2012){Sales}, {Navarro}, {Theuns}, {Schaye},
  {White}, {Frenk}, {Crain}, \& {Dalla Vecchia}}]{Sales12}
{Sales} L.~V., {Navarro} J.~F., {Theuns} T., {Schaye} J., {White} S.~D.~M.,
  {Frenk} C.~S., {Crain} R.~A., {Dalla Vecchia} C., 2012, \mnras, 423, 1544

\bibitem[{{Scannapieco} {et~al}\mbox{.}(2010){Scannapieco}, {Gadotti},
  {Jonsson}, \& {White}}]{S10}
{Scannapieco} C., {Gadotti} D.~A., {Jonsson} P., {White} S.~D.~M., 2010, MNRAS,
  407, L41

\bibitem[{{Scannapieco} {et~al}\mbox{.}(2005){Scannapieco}, {Tissera}, {White},
  \& {Springel}}]{S05}
{Scannapieco} C., {Tissera} P.~B., {White} S.~D.~M., {Springel} V., 2005,
  MNRAS, 364, 552

\bibitem[{{Scannapieco} {et~al}\mbox{.}(2006){Scannapieco}, {Tissera}, {White},
  \& {Springel}}]{S06}
{Scannapieco} C., {Tissera} P.~B., {White} S.~D.~M., {Springel} V., 2006,
  MNRAS, 371, 1125

\bibitem[{{Scannapieco} {et~al}\mbox{.}(2008){Scannapieco}, {Tissera}, {White},
  \& {Springel}}]{S08}
{Scannapieco} C., {Tissera} P.~B., {White} S.~D.~M., {Springel} V., 2008,
  MNRAS, 389, 1137

\bibitem[{{Scannapieco} {et~al}\mbox{.}(2009){Scannapieco}, {White},
  {Springel}, \& {Tissera}}]{S09}
{Scannapieco} C., {White} S.~D.~M., {Springel} V., {Tissera} P.~B., 2009,
  MNRAS, 396, 696 (S09)

\bibitem[{{Scannapieco} {et~al}\mbox{.}(2011){Scannapieco}, {White},
  {Springel}, \& {Tissera}}]{S11}
{Scannapieco} C., {White} S.~D.~M., {Springel} V., {Tissera} P.~B., 2011,
  MNRAS, 417, 154

\bibitem[{{Scannapieco} {et~al}\mbox{.}(2012){Scannapieco} {et~al.}}]{S12}
{Scannapieco} C., {et~al.}, 2012, \mnras, 423, 1726

\bibitem[{{Sellwood} \& {Binney}(2002)}]{Sellwood02}
{Sellwood} J.~A., {Binney} J.~J., 2002, \mnras, 336, 785

\bibitem[{{Smith} {et~al}\mbox{.}(2007){Smith}, {Ruchti}, {Helmi}, {Wyse},
  {Fulbright}, {Freeman}, {Navarro}, {Seabroke}, {Steinmetz}, {Williams},
  {Bienaym{\'e}}, {Binney}, {Bland-Hawthorn}, {Dehnen}, {Gibson}, {Gilmore},
  {Grebel}, {Munari}, {Parker}, {Scholz}, {Siebert}, {Watson}, \&
  {Zwitter}}]{Smith07}
{Smith} M.~C. {et~al.}, 2007, \mnras, 379, 755

\bibitem[{{Springel}(2005)}]{Springel05}
{Springel} V., 2005, MNRAS, 364, 1105

\bibitem[{{Springel} {et~al}\mbox{.}(2008){Springel}, {Wang}, {Vogelsberger},
  {Ludlow}, {Jenkins}, {Helmi}, {Navarro}, {Frenk}, \& {White}}]{Springel08}
{Springel} V. {et~al.}, 2008, MNRAS, 391, 1685

\bibitem[{{Springel} {et~al}\mbox{.}(2001){Springel}, {White}, {Tormen}, \&
  {Kauffmann}}]{Springel_2001}
{Springel} V., {White} S.~D.~M., {Tormen} G., {Kauffmann} G., 2001, MNRAS, 328,
  726

\bibitem[{{Steinmetz} \& {Navarro}(1999)}]{Steinmetz99}
{Steinmetz} M., {Navarro} J.~F., 1999, \apj, 513, 555

\bibitem[{{Stinson} {et~al}\mbox{.}(2013){Stinson}, {Bovy}, {Rix}, {Brook},
  {Ro{\v s}kar}, {Dalcanton}, {Macci{\`o}}, {Wadsley}, {Couchman}, \&
  {Quinn}}]{Stinson13}
{Stinson} G.~S. {et~al.}, 2013, \mnras, 436, 625

\bibitem[{{Trujillo-Gomez} {et~al}\mbox{.}(2015){Trujillo-Gomez}, {Klypin},
  {Col{\'{\i}}n}, {Ceverino}, {Arraki}, \& {Primack}}]{Trujillo15}
{Trujillo-Gomez} S., {Klypin} A., {Col{\'{\i}}n} P., {Ceverino} D., {Arraki}
  K.~S., {Primack} J., 2015, \mnras, 446, 1140

\bibitem[{{Vogelsberger} {et~al}\mbox{.}(2013){Vogelsberger}, {Genel},
  {Sijacki}, {Torrey}, {Springel}, \& {Hernquist}}]{Vogelsberger13}
{Vogelsberger} M., {Genel} S., {Sijacki} D., {Torrey} P., {Springel} V.,
  {Hernquist} L., 2013, \mnras, 436, 3031

\bibitem[{{Wang} {et~al}\mbox{.}(2012){Wang}, {Hammer}, {Athanassoula},
  {Puech}, {Yang}, \& {Flores}}]{Wang12}
{Wang} J., {Hammer} F., {Athanassoula} E., {Puech} M., {Yang} Y., {Flores} H.,
  2012, \aap, 538, A121

\bibitem[{{Watkins}, {Evans} \& {An}(2010){Watkins}, {Evans}, \&
  {An}}]{Watkins10}
{Watkins} L.~L., {Evans} N.~W., {An} J.~H., 2010, \mnras, 406, 264

\bibitem[{{Wilkinson} \& {Evans}(1999)}]{Wilkinson99}
{Wilkinson} M.~I., {Evans} N.~W., 1999, \mnras, 310, 645

\bibitem[{{Yepes}, {Gottl{\"o}ber} \& {Hoffman}(2014){Yepes}, {Gottl{\"o}ber},
  \& {Hoffman}}]{Yepes13}
{Yepes} G., {Gottl{\"o}ber} S., {Hoffman} Y., 2014, \nar, 58, 1

\bibitem[{{Ziparo} {et~al}\mbox{.}(2013){Ziparo} {et~al.}}]{Ziparo_2013}
{Ziparo} F., {et~al.}, 2013, \mnras, 434, 3089

\end{thebibliography}

\end{document}